\documentclass[journal]{IEEEtran}
\usepackage{amsmath,amsfonts}
\usepackage{amssymb}
\usepackage{algorithmic}
\usepackage{algorithm}
\usepackage{array}
\usepackage{bm}
\usepackage{booktabs}
\usepackage{cite}
\usepackage{color}
\usepackage{esint}
\usepackage[mathscr]{eucal}
\usepackage{makecell}
\usepackage{multicol}
\usepackage[caption=false,font=small,labelfont=rm,textfont=rm]{subfig}
\usepackage{stfloats}
\usepackage{subfig,graphicx}
\usepackage{textcomp}
\usepackage{url}
\usepackage{verbatim}
\makeatletter

\newcommand{\Rmnum}[1]{\expandafter\@slowromancap\romannumeral #1@}
\makeatother

\begin{document}

\title{A Pixel-based Reconfigurable Antenna Design for \\ Fluid Antenna Systems}

\author{Jichen Zhang,~\IEEEmembership{Graduate Student Member,~IEEE,} Junhui Rao,~\IEEEmembership{Graduate Student Member,~IEEE,} \\Zhaoyang Ming,~\IEEEmembership{Graduate Student Member,~IEEE,} Zan Li,~\IEEEmembership{Graduate Student Member,~IEEE,} \\Chi-Yuk Chiu,~\IEEEmembership{Senior Member,~IEEE,} Kai-Kit Wong,~\IEEEmembership{Fellow,~IEEE,} Kin-Fai Tong,~\IEEEmembership{Fellow,~IEEE,} \\Ross Murch,~\IEEEmembership{Fellow,~IEEE}
\thanks{This work was supported by the Hong Kong Research Grants Council Collaborative Research Fund under Grant AoE/E-601/22-R. (\it{Corresponding author: Junhui Rao.})}
\thanks{Jichen Zhang, Junhui Rao, Zhaoyang Ming, Zan Li, and Chi-Yuk Chiu are with the Department of Electronic and Computer Engineering, the Hong Kong University of Science and Technology, Hong Kong (e-mail: jzhangiq@connect.ust.hk).}
\thanks{Kai-Kit Wong and Kin-Fai Tong are with the Department of Electronic and Electrical Engineering, University College London, London WC1E 7JE, U.K (e-mail: kitwong@ieee.org, k.tong@ucl.ac.uk).}
\thanks{Ross Murch is with the Department of Electronic and Computer Engineering and the Institute for Advanced Study (IAS), the Hong Kong University of Science and Technology, Hong Kong (e-mail: eermurch@ust.hk).}
}



\maketitle

\begin{abstract}
Fluid Antenna Systems (FASs) have recently been proposed for enhancing the performance of wireless communication. Previous antenna designs to meet the requirements of FAS have been based on mechanically movable or liquid antennas and therefore have limited reconfiguration speeds. In this paper, we propose a design for a pixel-based reconfigurable antenna (PRA) that meets the requirements of FAS and the required switching speed. It can provide 12 FAS ports across 1/2 wavelength and consists of an E-slot patch antenna and an upper reconfigurable pixel layer with 6 RF switches. Simulation and experimental results from a prototype operating at 2.5 GHz demonstrate that the design can meet the requirements of FAS including port correlation with matched impedance. 
\end{abstract}


\section{Introduction}

By 2029 it is expected that total mobile data traffic (excluding traffic generated by fixed wireless access) will grow to 403 EB per month, from 130 EB per month at the end of 2023  \cite{ericsson}. To meet this demand, the development of the next generation of wireless communication, the 6-th generation (6G), has already begun \cite{6G_1,6G_2,6G_3}. It is thought that 6G will need to rely on a suite of technologies that span fundamental electromagnetic structures such as multiple-input multiple-output (MIMO) and reconfigurable intelligent surfaces (RISs) through to artificial intelligence (AI), all of which are currently being investigated \cite{6G_HMIMO,6G_MIMO,6G_RIS}. The development of 6G will also require new technologies to be developed and one technology that has been shown to have potential for use in 6G is the Fluid Antenna System (FAS) \cite{FAS,BruceLee}. FAS promises to enhance wireless system performance and also reduce implementation cost \cite{history_of_FAS_MA}.

\begin{figure}[!t]
	\centering
	\subfloat[]{\includegraphics[width=0.95\linewidth]{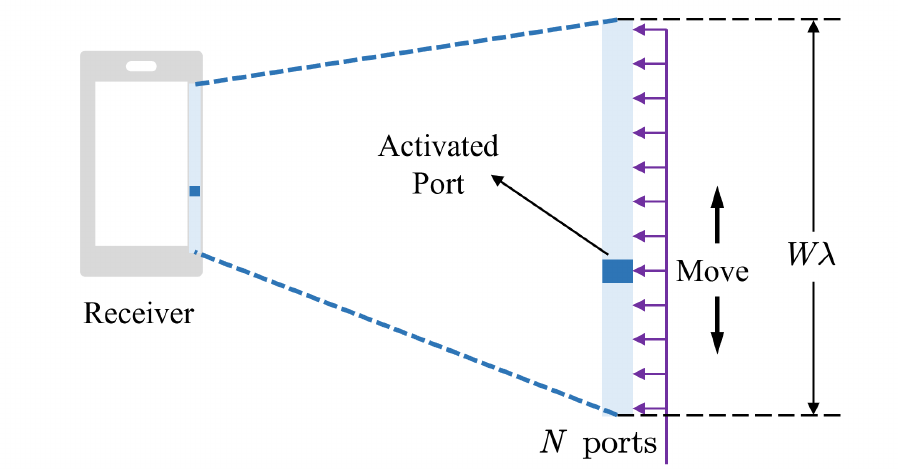}}
	\hfil
	\subfloat[]{\includegraphics[width=1\linewidth]{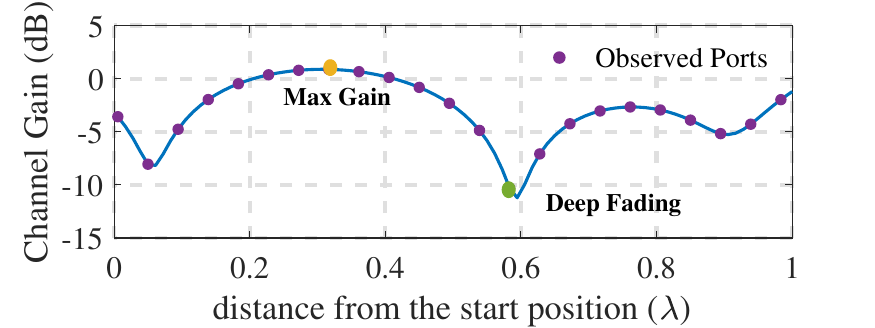}}
	\hfil
	\subfloat[]{\includegraphics[width=1\linewidth]{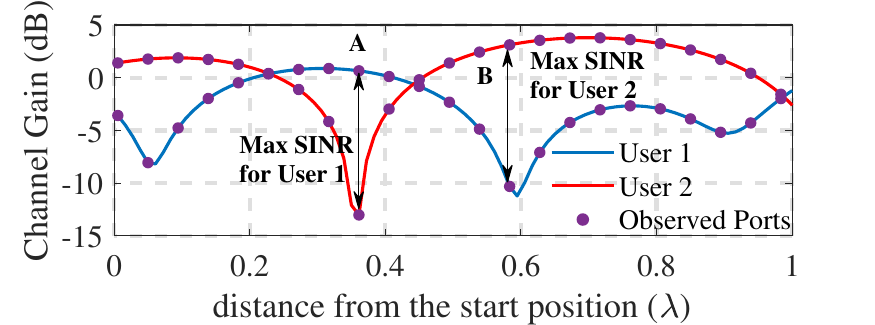}}
	\hfil
	\caption{(a) Systems impression of a linear FAS with size $W\lambda$ and $N$ ports. (b) Channel gain for different ports of FAS. (c) Channel gain at different ports of FAS with two base stations. Port position A achieves maximum SIR for User 1, while port position B achieves maximum SIR for User 2.}
	\label{FAS_Scheme}
\end{figure}

FAS, in its most straightforward form, can be thought of as a single antenna element that has $N$ preset port locations (known as FAS ports) that are evenly located over a certain length $W\lambda$ (where $\lambda$ is wavelength). The FAS antenna can be thought of as a single radiating element that is physically moved to provide the $N$ port positions. A systems impression of its implementation \cite{FAS,BruceLee} is shown in Fig. \ref{FAS_Scheme}(a). At any instant in time, only one of the FAS ports can be accessed. In a rich scattering environment, the channel gain at each FAS port will follow a Rayleigh distribution that is spatially correlated. Because the ratio $N/W$ is $\gg 1$, the ports of the FAS can finely sample the Rayleigh fading spatial signal as shown by the solid dots in Fig. \ref{FAS_Scheme}(b). Due to the fine spatial sampling, spatial channel correlation between adjacent ports of the FAS is evident, and it is this channel correlation feature that FAS exploits to enhance wireless communication performance. 

One example of how fine spatial sampling of the channel can be exploited is shown in Fig. \ref{FAS_Scheme}(c), where we have a multi-user rich scattering environment in which signals from two base stations are received by the FAS \cite{f-FAMA}. Since the base stations are far apart, the signal fading seen at the FAS will be different for each base station. Therefore, the FAS can select the spatial port where the signal from the desired base station is high and the signal from the interfering one is low, as shown by point A in Fig. \ref{FAS_Scheme}(c). That is, by finely selecting the FAS port position, we can significantly increase the signal-to-interference ratio (SIR) to enhance communication performance. More generally, it has been shown that using FAS for multiple access, known as Fluid Antenna Multiple Access (FAMA), can approach near optimal performance for two users \cite{s-FAMA,Achievability_FAMA}. Furthermore, FAMA can be utilized in conjunction with other communication technologies, including millimeter-wave (mmW) communication, RIS, MIMO, and non-orthogonal multiple access (NOMA) \cite{BruceLee,MMW_FAS,CUMA,FAS_Part2,FAMA_NOMA,FAMA_NOMA_short}. FAMA is just one possible example of how FAS may be utilized and there are other scenarios also under investigation.

The development of FAS was inspired from the wireless systems perspective where there are numerous results. There are fewer examples of FAS antenna designs and these have been primarily based on mechanical antennas including liquid-based \cite{liquid1,liquid2}, surface-wave-based \cite{surface_wave}, and programmable-droplet-based \cite{droplet}. They can largely meet the system specifications of FAS by moving metal or liquid in the antenna to achieve fine spatial sampling. However, because the designs rely on physical movement, the spatial sampling can only occur relatively slowly, which can significantly constrain the FAS performance \cite{s-FAMA}. Compared to the packet transmission rate (around one millisecond per packet), existing mechanical-based antenna designs are not fast enough \cite{BruceLee} to provide the packet-by-packet reconfiguration required by FAS. 

In this work, we propose a new approach to FAS antenna design that leverages pixel-based reconfigurable antenna (PRA) design. Because the PRAs use electronic switching, such as PIN diodes, they can be configured at microsecond speeds and therefore meet the needs of packet-to-packet reconfigurability required in FAS. PRA has a long history of development and dates back to 2004 when they were first proposed \cite{PRA_start}. Since then a large number of designs have been proposed that can reconfigure pattern, polarization, and frequency \cite{pixel0,MIMO_pixel,pixel_RIS,pixel1,pixel2,pixel3,pixel_slot,RA,IMPM}.

One of the challenges in using PRA for FAS antennas is that previous PRAs have not been designed to provide fine spatial sampling of the channel. Therefore, in this paper, we provide a PRA design that addresses this issue and meets the requirements of FAS. In the remainder of this paper, we refer to PRA designs that meet the requirements of FAS as PRA-FAS. The key research contributions and distinctive features of our proposed design are as follows:

\begin{itemize}
	\item [1)]
		{\it{Pattern Covariance Matrix Analysis}}: We relate signal correlation to the radiation patterns of the antennas and show that FAS can also be interpreted as a $''$fluid$''$ radiation pattern. Using this approach, a radiation pattern covariance matrix is introduced to describe the radiation pattern correlation (and therefore signal correlation) between any two ports of the PRA-FAS. In addition, an analytical method for the evaluation of the covariance among PRA-FAS ports is also provided.
	\item [2)]
		{\it{FAS Ports}}: The proposed PRA-FAS design achieves reconfigurability, achieving 12 FAS ports within a compact volume of $0.67\lambda \times 0.67\lambda \times 0.125\lambda$. This level of port density can match that of traditional fluid antennas.
	\item [3)]
		{\it{Reconfigurability Speed}}: By leveraging RF switches and a software-controllable mechanism, the proposed PRA-FAS achieves port switching with $\mathrm{\mu s}$ latency, meeting the switching requirements of FAS \cite{f-FAMA,s-FAMA}.
	\item [4)]
		{\it{Two-step Method for PRA-FAS Design}}: A random search and Genetic Algorithm (GA) is devised to optimize the PRA-FAS design. The two-step method is necessary in order to reduce the design complexity of the PRA-FAS design. First, pixel configurations that provide a good antenna match are found. Then from these configurations, those that meet the FAS spatial correlation are selected.  
	\item [5)] 
		{\it{PRA-FAS Prototype and Experimental Results}}: We provide a prototype design that operates in the 2.5 GHz band. Experimental and simulation results of the design are used to demonstrate its operation as well as providing the necessary spatial correlation requirements.
\end{itemize}

The remainder of the paper is organized as follows. Section \Rmnum{2} describes the system model of FAS and the design objective for PRA-FAS. Section \Rmnum{3} provides the geometry of the proposed PRA-FAS and the model for its analysis. In Section \Rmnum{4} we utilize the model to obtain an optimum design for the PRA-FAS and introduce an efficient two-step searching algorithm. The performance of the proposed antenna is reported by simulation and measurement results of the prototype in Section \Rmnum{5}. Section \Rmnum{6} provides a discussion of some key PRA-FAS issues. Finally, Section \Rmnum{7} provides conclusions. 

\textit{Notation}: For brevity, we leave out the explicit dependence of frequency on electromagnetic quantities such as impedance and radiation patterns, unless otherwise stated. Letters in bold font denote matrices or vectors, while letters not in bold font represent scalars. $\theta$ and $\phi$ represent spherical coordinates and are termed the elevation and azimuth angles in this work, with $\hat{\bm{\theta}}$ and $\hat{\bm{\phi}}$ being the unit spherical coordinate vectors. $[\mathbf{A}]_i$ is the $i$-th element of the vector $\mathbf{A}$, while $[\mathbf{A}]_{i,j}, \mathbf{A}^\mathrm{T}$, and $\mathbf{A}^\mathrm{H}$ are the $(i,j)$-th entry, the transpose, and the conjugate transpose of the matrix $\mathbf{A}$, respectively. Letters in math calligraphy font denote the set. The cardinal number of set $\mathcal{U}$ is denoted as $\mathrm{card}(\mathcal{U})$. $\mathbb{C}$ refers to the complex number set.  $\cal{E}\left[\cdot\right]$ denotes expected value. $\delta(\cdot)$ denotes the impulse function. $\mathbf{U}_N$ is the $N \times N$ identity matrix. $\mathrm{diag}(a_1,a_2,\dots,a_N)$ denotes a diagonal matrix with diagonal elements $a_1,a_2,\dots,a_N$, while $\mathrm{diag}(\mathbf{A}_1,\mathbf{A}_2,\dots,\mathbf{A}_N)$ denotes the block diagonal matrix with diagonal elements $\mathbf{A}_1,\mathbf{A}_2,\dots,\mathbf{A}_N$.

\section{PRA-FAS Design Objective}

The FAS illustrated in Fig. \ref{FAS_Scheme}(a) utilizes a traversable or shape-flexible radiator capable of switching between $N$ port positions uniformly distributed along a linear space of $W\lambda$. The FAS antenna can be thought of as a single radiating element that is physically moved to provide the $N$ port positions. We consider the case where only one port at a time can be activated due to the necessary physical movement. 

To analyze the antenna, let the open circuit voltage received at the $n$-th FAS receiver port due to the base-station transmitter be denoted $g_n$. 
The received open circuit voltages of all $N$ ports are written as a vector $\bm{g} = \left[ g_1,g_2,\dots,g_N \right]^\mathrm{T}$.

To characterize the fine spatial sampling in FAS, it is necessary to obtain a statistical relation between the $g_n$. As such a FAS channel model describing the correlation relationship between the various ports in FAS has been proposed \cite{FAS_new_model}. This is performed by adopting a spatial covariance matrix  $\bm{\varrho}$ between the $(i,j)$-th ports and it is typically specified as 
\begin{equation}
	\label{Covariance_ij}
	\left[ \bm{\varrho} \right]_{i,j} = \rho_{i,j}= \mathrm{Cov} ( g_i, g_j ) =J_0\left(\frac{2\pi|i-j|W}{N-1}\right),
\end{equation}
where $J_0$ is the Bessel function of the first kind, order zero.  It can be seen as following Clarke's model for mobile radio propagation 
\cite{Clarke_Model}.
 
To relate the requirement (\ref{Covariance_ij}) to antenna design, it is necessary to define the electromagnetic environment. Let the incident radiation at the FAS be written as
\begin{equation}
	\label{eqn-incident}
	\mathbf{h}(\mathbf{\Omega}) 
	= h_\theta(\mathbf{\Omega})\hat{\bm{\theta}}+h_\phi(\mathbf{\Omega})\hat{\bm{\phi}}
	= [h_\theta(\mathbf{\Omega}),h_\phi(\mathbf{\Omega})]^\mathrm{T},
\end{equation} 
where $\mathbf{\Omega} = (\theta, \phi)$. The corresponding polarization matrix is then written as \cite{1969Antenna, TVT}
\begin{equation}
	\label{S_angular}
	\bm{\Gamma}(\mathbf{\Omega,\Omega'})=\begin{bmatrix} \varGamma_{\theta,\theta}(\mathbf{\Omega},\mathbf{\Omega}'), & \varGamma_{\theta,\phi}(\mathbf{\Omega},\mathbf{\Omega}') \\ \varGamma_{\phi,\theta}(\mathbf{\Omega},\mathbf{\Omega}'), & \varGamma_{\phi,\phi}(\mathbf{\Omega},\mathbf{\Omega}') \end{bmatrix} \in \mathbb{C}^{2 \times 2},
\end{equation}
where $\varGamma_{\theta,\phi}(\mathbf{\Omega},\mathbf{\Omega}')= \mathcal{E} \left[ h^{}_\theta(\mathbf{\Omega}) h^{*}_\phi(\mathbf{\Omega}')\right]$, and similarly for  $\varGamma_{\theta,\theta}(\mathbf{\Omega},\mathbf{\Omega}')$,  $\varGamma_{\phi,\theta}(\mathbf{\Omega},\mathbf{\Omega}')$ and  $\varGamma_{\phi,\phi}(\mathbf{\Omega},\mathbf{\Omega}')$.

In rich scattering scenarios, the incident radiation is modeled with both polarization components uncorrelated and equal in power, so that $\varGamma_{\phi,\phi}=\varGamma_{\theta,\theta}$ and $\varGamma_{\phi,\theta}$ and $\varGamma_{\theta,\phi}$ are zero and $\bm{\Gamma}(\mathbf{\Omega},\mathbf{\Omega}')$ is a diagonal matrix. In addition, we can assume that the spatial components are also uncorrelated and write the resulting polarization matrix as
\begin{equation}
	\label{eqn-PAS} 
	\bm{\Gamma}(\mathbf{\Omega},\mathbf{\Omega}')=S(\mathbf{\Omega})\mathbf{U}_2 \delta(\mathbf{\Omega}-\mathbf{\Omega}'),
\end{equation}
where $S(\mathbf{\Omega})$ is known as the power angular spectrum (PAS) and $\mathbf{U}_2$ is the $2 \times 2$ identify matrix. 

By using (\ref{eqn-PAS}), we can obtain a statistical relation between $g_n$ at different ports by using correlation. The correlation coefficient for the open circuit voltages $g_i$ and $g_j$ of ports $i$ and $j$ in the presence of the rich scattering scenario can be written as 
\begin{equation}
	\label{eqn-oc}
	\rho_{i,j}=\frac{\mathcal{E}\left[g_i^{}g_j^*\right]}
	{\sqrt{\mathcal{E}\left[g_i^{} g_i^*\right]\mathcal{E}\left[g_j^{} g_j^*\right]}},
\end{equation}
where it is assumed the expected values $\mathcal{E}\left[g_n\right]$ are all zero. The voltages at the ports can be expressed in terms of the incident radiation and the FAS pattern of the $n$-th port as \cite{1969Antenna,TVT}
\begin{equation}
	\label{eqn-height}
	g_n=a\iint_{\mathbf{\Omega}}  \mathbf{e}_n(\mathbf{\Omega}) \cdot \mathbf{h}_n(\mathbf{\Omega}) \mathrm{d} \mathbf{\Omega} ,
\end{equation} 
where $\mathbf{e}_n(\mathbf{\Omega}) = e_{\theta ,n}(\mathbf{\Omega})\hat{\bm{\theta}}+e_{\phi ,n}(\mathbf{\Omega})\hat{\bm{\phi}} = \left[e_{\theta, n}(\mathbf{\Omega}),e_{\phi,n}(\mathbf{\Omega})\right]^\mathrm{T}$ is the FAS radiation pattern of the $n$-th port. The proportionality constant $a$ is  given in \cite{1969Antenna,Harvesting} but is not required as it will be canceled later. Using (\ref{eqn-oc}) and (\ref{eqn-height}), the expression for correlation can be written as 
\begin{equation}
	\label{eqn-corr}
	\rho_{i,j}=\mathcal{E}\left[\iint_{\mathbf{\Omega}_i} \mathbf{e}_i(\mathbf{\Omega}_i) \cdot \mathbf{h}_i(\mathbf{\Omega}_i) \mathrm{d} \mathbf{\Omega}_i  
	\iint_{\mathbf{\Omega}_j} \mathbf{h}^*_j(\mathbf{\Omega}_j) \cdot \mathbf{e}^*_j(\mathbf{\Omega}_j) \mathrm{d}\mathbf{\Omega}_j \right],
\end{equation}
where the denominator terms in (\ref{eqn-oc}) have not been included for expediency. Imposing the rich scattering scenario \cite{Clarke_Model,TVT}, using (\ref{eqn-PAS}) and interchanging the order of integration and expectation, we arrive at the final expression for correlation between ports as 
\begin{equation}
	\label{Pattern_Correlation}
	\rho_{i,j}=
	\frac{\iint \mathbf{e}^{}_i(\mathbf{\Omega}) \cdot \mathbf{e}_j^*(\mathbf{\Omega}) S\mathbf{(\Omega)} \mathrm{d}\mathbf{\Omega}}
	{\sqrt{\iint \mathbf{e}_i^{}(\mathbf{\Omega}) \cdot \mathbf{e}_i^*(\mathbf{\Omega}) S\mathbf{(\Omega)}  \mathrm{d} \mathbf{\Omega} \iint \mathbf{e}_j^{}(\mathbf{\Omega}) \cdot \mathbf{e}_j^*(\mathbf{\Omega}) S\mathbf{(\Omega)}  \mathrm{d} \mathbf{\Omega}}}.
\end{equation}

Equation (\ref{Pattern_Correlation}) implies that the correlation characteristics between two FAS ports only depend on the patterns of the antennas and the scattering environment through its PAS. It also suggests another interpretation of FAS that can capture the definition of the shape and position flexible antennas, i.e., the fluid antenna can be thought of as equivalent to a fluid radiation pattern. In particular, the radiation patterns of the different FAS ports are transformed or configured from one to another to meet the required port FAS correlation (\ref{Covariance_ij}).
 
To provide an example of the equivalence between shape and position flexible antennas to $''$fluid$''$ radiation patterns, we can consider the canonical example of two vertically polarized dipoles where incident radiation is restricted to the 2-dimensional (2D) plane so that $S(\mathbf{\Omega})=\delta(\theta)$. That is we assume all the ports of the FAS act as vertically polarized dipoles but are separated by distance $d_{i,j}=\frac{|i-j|W}{N-1}$ in the horizontal plane. If the pattern of the $i$-th port is defined as $\mathbf{e}_i(\mathbf{\Omega})=E(\theta)\hat{\bm{\theta}}$, then that of the $j$-th port can be obtained by phase translation as  $\mathbf{e}_j(\mathbf{\Omega})=E(\theta)e^{jkd_{i,j}\cos \phi}\hat{\bm{\theta}} $. Using (\ref{Pattern_Correlation}) we can find the correlation as 
\begin{eqnarray}
	\label{eqn-dipole}
	\begin{split}
	\rho_{i,j}&=\iint E(\theta) E^*(\theta) e^{-jkd_{i,j}\cos \phi} \delta(\theta) \mathrm{d} \mathbf{\Omega} \\
	&= J_0\left(\frac{2\pi|i-j|W}{N-1}\right),
	\end{split}
\end{eqnarray}
where we have not included the denominator terms from (\ref{Pattern_Correlation}) in the first line for expediency. As can be observed we obtain exactly the same result as in (\ref{Covariance_ij}) showing that changing the FAS radiation pattern can achieve the same results as moving the dipole antenna when in the rich scattering environment. 


Equation (\ref{Pattern_Correlation}) and the example  (\ref{eqn-dipole}), show that the patterns can be utilized to obtain port correlations and these depend on the particular scattering environment of PAS that is utilized. More generally, (\ref{Pattern_Correlation}) is valid for FAS ports with arbitrary antenna patterns.  

Using (\ref{Pattern_Correlation}) and the example  (\ref{eqn-dipole}), the general design objective of PRA-FAS can now be defined. Our objective is to design an antenna with $N$ reconfigurable states such that the elements of formula (\ref{Pattern_Correlation}) are the same as specified by the required FAS covariance (\ref{Covariance_ij}) for a given $W$. A subtlety of the design objective is that the selection of the states must also be ordered appropriately to meet (\ref{Covariance_ij}). From a systems perspective, meeting the correlation requirement has not been proven to be optimal and other correlation functions may later be found to be better. Nevertheless, for now, our objective is to meet the conditions imposed by (\ref{Covariance_ij}). However, it is also worth noting that a significant advantage of our design approach is that it is general and can potentially match any required FAS covariance. 
 
In this work, the specific objective is for a PRA-FAS with $N=12$ and $W=1/2$. This implies that we need to design an antenna with $N=12$ patterns that fulfill the conditions imposed by (\ref{Covariance_ij}) with $W=1/2$ using (\ref{Pattern_Correlation}). In the following sections, we describe our approach to meeting the specific design objective we have specified.  

\section{Proposed PRA-FAS Geometry and Model}

\begin{figure}[!t]
	\centering
	\includegraphics[width=1\linewidth]{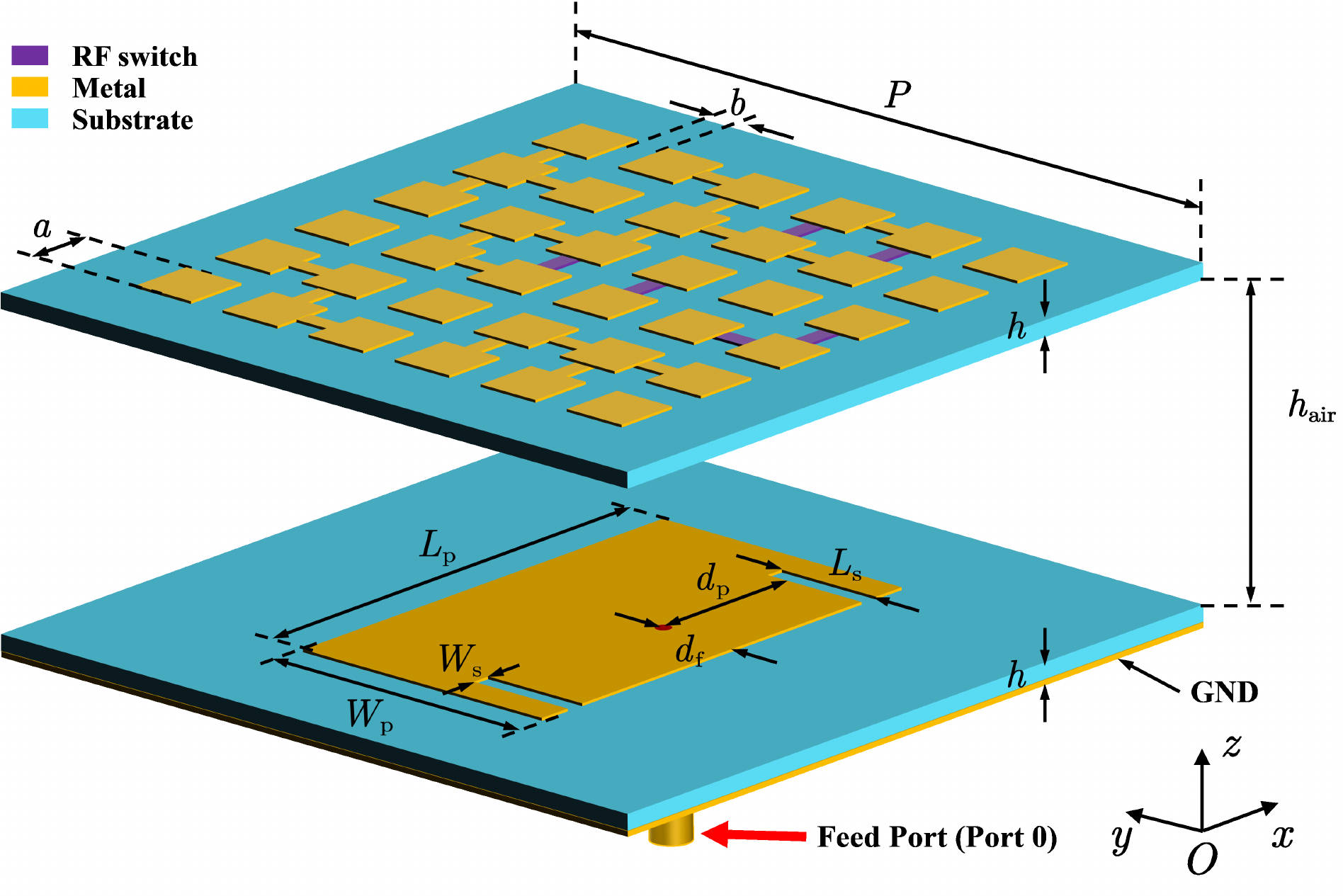}
	\caption{Configuration of the proposed PRA-FAS. Dimensions: $P=80$, $a=7$, $b=4$, $h=1.524$, $h_\mathrm{air}=12$, $L_\mathrm{p}=50$, $W_\mathrm{p}=30.6$, $W_\mathrm{s}=2$, $L_\mathrm{s}=12$, $d_\mathrm{p}=19.4$, and $d_\mathrm{f}=10.7$ (all in millimeters).}
	\label{Antenna_3D}
\end{figure}

\begin{figure}[!t]
	\centering
	\includegraphics[width=0.9\linewidth]{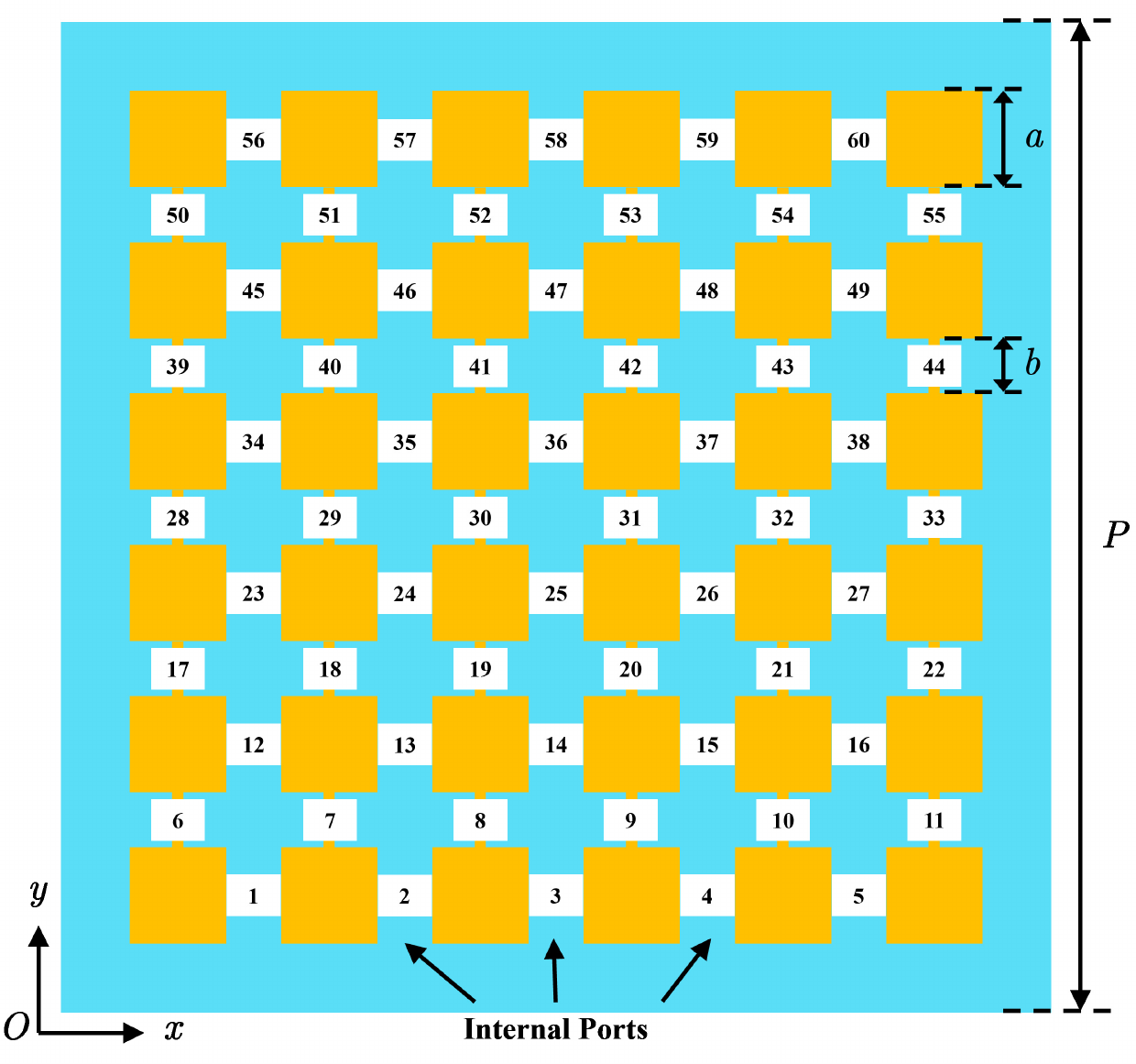}
	\caption{Configuration of the upper pixel layer, which includes $Q=60$ potential connections between adjacent pixels. The numbering of potential connections is also shown, and we define these connections as internal ports as discussed later.}
	\label{Pixel_layer}
\end{figure}

To meet the design objective, we propose the PRA-FAS geometry as shown in Fig. \ref{Antenna_3D}. It consists of an E-slot patch that is probe-fed through its ground plane as shown. The upper pixel substrate consists of metallic pixels printed on its top layer. The E-slot patch acts as a radiation source for the upper pixel substrate. Different radiation characteristics can be obtained by utilizing different configurations for connections between the pixels on the top layer. There are $Q=60$ possible positions for connections between adjacent pixels and this can be seen with reference to Fig. \ref{Pixel_layer}. Both the upper and lower substrates  consist of Rogers 4003C material with a permittivity of $\epsilon_r = 3.55$ and a loss tangent of $\tan \delta_d = 0.0027$. 

The reason why the structure in Fig. \ref{Antenna_3D} has been proposed is that its input impedance is generally well-matched for a wide range of pixel connection configurations on the upper substrate. From this set of connection configurations, we can then select or search through the radiation patterns to find 12 that meet the required spatial correlation design objective, namely (\ref{Covariance_ij}) using (\ref{Pattern_Correlation}) in this work. 

For an ideal PRA scenario, all $Q$ connections between pixels can be controlled by RF switches, allowing up to $2^Q$ configurations (ignoring possible symmetries) and therefore $2^Q$ patterns. However, controlling a large number of RF switches would necessitate a complex DC feeding network \cite{pixel0}, compromising the performance of the antenna in terms of efficiency \cite{complexity}. To mitigate this complexity, we deploy a limited number of $P (<Q)$ RF switches and fix the other connections between pixels with hardwires or open circuits. As indicated in Fig. \ref{Antenna_3D}, there are $P=6$ switches between pixels in our proposed design. The number of switches $P=6$ has been selected as a good tradeoff between design flexibility and complexity (see the discussion section later where we provide more details on how to select the number of switches). With $P=6$, this can provide at most $2^P=64$ patterns, from which we must find 12 that are well matched and meet the spatial correlation requirements in (\ref{Covariance_ij}). 



To capture the configuration of the PRA-FAS with $P$ RF switches and $Q$ potential connections, the states of the connections should be specified. As a result, a binary variable $x_q \in \{0,1\}$ for $q = 1,2,\dots, Q$ is utilized to express whether the $q$-th connection is open ($''$0$''$) or connected ($''$1$''$). The states of the $Q$ connections can then be written as a vector
\begin{equation}
	\label{fixed_bits}
	\mathbf{x} = \left[ x_1, x_2, \dots, x_Q \right].
\end{equation}
Because the connections between pixels can be either switches or fixed, it is also necessary to differentiate them in (\ref{fixed_bits}). This is required later because the impedance of the switches are different from the hardwires, and therefore need to be specified. The most direct way to perform this is to specify which connections are switches. This can be performed by specifying a set with the connection numbers of those with switches. For the $P$ switches, these positions can be given by $P$ connection numbers and specified by the set ${\cal S}$ as
\begin{equation}
	\label{switch_ports}
	{\cal S} = \{ q_1, q_2, \dots, q_P \} \ \mathrm{for} \ P<Q,
\end{equation}
where the elements $q_1$ to $q_P$ indicate the connection numbers with switches. The vectors $\mathbf{x}$ and the set ${\cal S} $ then completely specify the configuration of the pixel layer in the PRA-FAS. 

\begin{figure}[!t]
	\centering
	\includegraphics[width=0.9\linewidth]{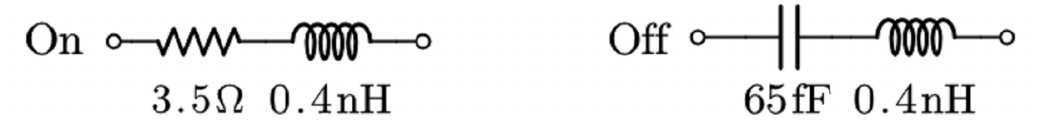}
	\caption{Equivalent circuit model of the RF switch MA4AGP907.}
	\label{Circuit_of_switch}
\end{figure}

To model our PRA-FAS we first express the connections between adjacent pixels in the FAS-PRA using impedance. For a particular $\mathbf{x}$ and ${\cal S}$, we use the diagonal $Q \times Q$ impedance matrix $\mathbf{Z^L}=\mathrm{diag}\left(Z^\mathrm{L}_1,Z^\mathrm{L}_2,\dots,Z^\mathrm{L}_Q\right)$ to represent the connections. The off-diagonal elements are all zero because there are only connections between adjacent pixels.  For those connections with $q \notin \cal{ S}$ (that is short by hardwires or open), we straightforwardly set the impedance matrix element $Z^\mathrm{L}_q$ to zero or $\infty$ depending on whether $x_q$  is 0 or 1, respectively. For the six connections with switches, as specified by ${\cal S}$, we need to use the on and off impedance of the RF switches. In our work, we will use the RF PIN diode MA4AGP907 \cite{datasheet907} and its equivalent circuit models for the on and off states are given in Fig. \ref{Circuit_of_switch}. Therefore, for those connections $q \in {\cal S}$, we set the element $Z^\mathrm{L}_q$ to be either the on or off impedance shown in  Fig. \ref{Circuit_of_switch}, specified by whether the corresponding element $x_q$ is 0 or 1, respectively.  

\begin{figure}[!t]
	\centering
	\includegraphics[width=0.9\linewidth]{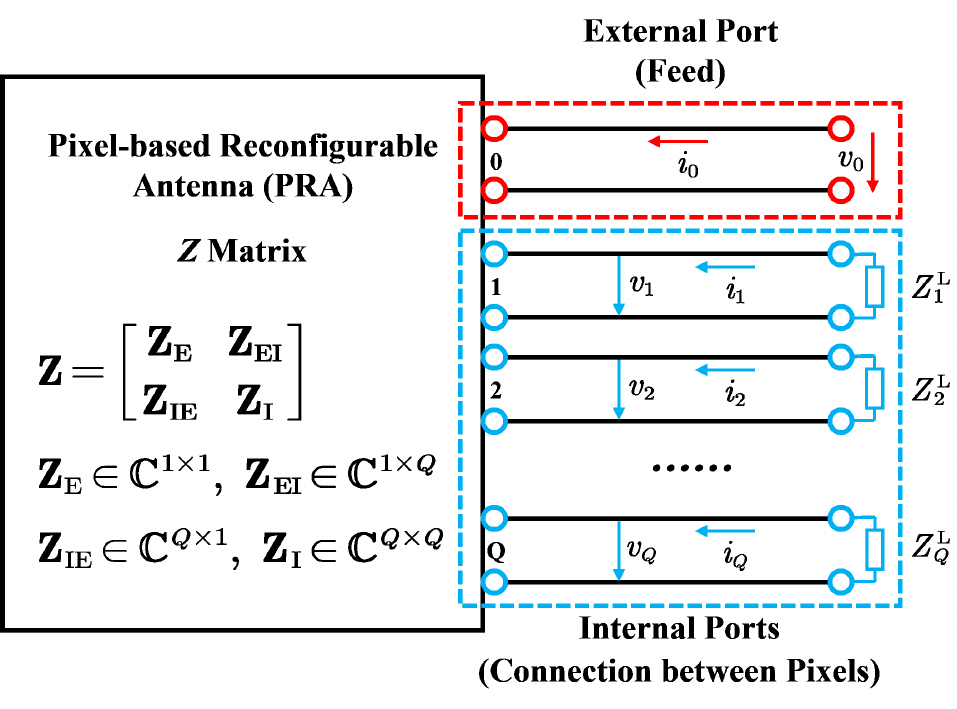}
	\caption{Circuit model for the PRA-FAS with one feeding port and $Q$ internal ports.}
	\label{IMPM}
\end{figure}

With the impedance of the pixel connections specified, we can now provide a circuit model for the entire PRA-FAS as shown in Fig. \ref{IMPM}. The technique that utilizes this model has been previously termed the internal multi-port method (IMPM) \cite{IMPM}. In IMPM, we can represent the FAS-PRA using $(Q+1)$ ports made up of $Q$ internal ports (representing the $Q$ load impedance connections) and one external port (the single external feed port). The model is accurate as long as the coupling between the loads is negligible which is usually valid \cite{IMPM}. We number these ports with $q=0,1,2,\dots,Q$ (where 0 is the external feed port and $1$ to $Q$ are the internal ports). The voltage across the $q$-th port is denoted as $v_q$ and it has been associated with its current $i_q$. We group all $v_q$ and $i_q$ into vectors as
\begin{equation}
	\begin{split}
		\mathbf{v}&=\left[v_0,v_1,v_2,\dots,v_Q\right]^\mathrm{T}, \\
		\mathbf{i}&=\left[i_0, i_1, i_2,\ \dots,i_Q\right]^\mathrm{T},
	\end{split}
\end{equation}
where $v_0, i_0$ are the voltage and the current of the external feeding (port 0), and we define $\mathbf{v}^\mathbf{I}=\left[v_1,v_2,\dots,v_Q\right]^\mathrm{T},\mathbf{i}^\mathbf{I}=\left[i_1, i_2,\ \dots,i_Q\right]^\mathrm{T}$ as the voltage and current vector of the $Q$ internal ports, respectively. 

The impedance of the internal ports and external port is denoted by $\mathbf{Z}$, a $(Q+1)\times (Q+1)$ matrix so that the voltages and currents on the internal ports and external port have the following relationship
\begin{equation}
	\label{v_i}
	\mathbf{v}=
	\begin{bmatrix}
		v_0 \\ \mathbf{v}^\mathbf{I}
	\end{bmatrix}
	= \mathbf{Z} \mathbf{i} = 
	\begin{bmatrix}
		\mathbf{Z_E} \ &\mathbf{Z_{EI}} \\
		\mathbf{Z_{IE}} \ &\mathbf{Z_{I}}
	\end{bmatrix}
	\begin{bmatrix}
		i_0 \\ \mathbf{i}^\mathbf{I}
	\end{bmatrix},
\end{equation}
where $\mathbf{Z_E},\mathbf{Z_{EI}},\mathbf{Z_{IE}},\mathbf{Z_I}$ are shown in Fig. \ref{IMPM}. The voltage and current on the $q$-th internal port can be derived using the corresponding load impedance as $v_q = -Z^\mathrm{L}_q i_q$. Thus, the voltages and currents on all internal ports can be expressed as
\begin{equation}
	\mathbf{v^I}=-\mathbf{Z^L}\mathbf{i^I}.
\end{equation}

With a certain connection vector $\mathbf{x}$ and set $\mathcal{S}$, the input impedance of the PRA-FAS can be calculated as
\begin{equation}
	\label{Zin}
	\vspace{-0.1cm}
	Z_\mathrm{in}(\mathbf{x},\mathcal{S}) = \mathbf{Z_E}-\mathbf{Z_{EI}} \left[ \mathbf{Z_I} + \mathbf{Z^L}(\mathbf{x},\mathcal{S}) \right]^{-1} \mathbf{Z_{IE}},
\end{equation}
where $\mathbf{Z^L}(\mathbf{x},\mathcal{S})$ is the corresponding diagonal matrix, indicating the impedance terminated at all internal ports, either short, open, or switch in on/off states under a certain configuration. 

The subsequent radiation pattern of the PRA-FAS can also be found straightforwardly. With the port current vector, the open-circuit radiation patterns induced by the currents at each port can be summed together to obtain the total radiation pattern. Assuming that among the $N$ operating states in the PRA-FAS, the $n$-th radiation pattern, $\mathbf{e}_n(\mathbf{\Omega})$, excited by the $n$-th current vector, $\mathbf{i}_n$, is
\begin{equation}
	\vspace{-0.3cm}
	\label{total_Pattern}
	\mathbf{e}_n\left(\mathbf{\Omega}\right)=\sum_{q=0}^{Q}\left[\mathbf{i}_n\right]_q \mathbf{e}_q^{\mathrm{oc}}\left(\mathbf{\Omega}\right)=\mathbf{E}_{\mathrm{oc}}\mathbf{i}_n,
\end{equation}
where $\mathbf{e}^{\mathrm{oc}}_{i}(\mathbf{\Omega}) =\left[e^{\mathrm{oc}}_{\theta,i}(\mathbf{\Omega}), e^{\mathrm{oc}}_{\phi,i}(\mathbf{\Omega}) \right]^\mathrm{T}$ is the open-circuit radiation pattern excited by a unit current at the $q$-th port when all other ports are open, and $\mathbf{E}_{\mathrm{oc}}=\left[\mathbf{e}_0^{\mathrm{oc}},\mathbf{e}_1^{\mathrm{oc}},\mathbf{e}_2^{\mathrm{oc}},\cdot\cdot\cdot,\mathbf{e}_Q^{\mathrm{oc}}\right]$ denotes the combination of $\mathbf{e}_q^{\mathrm{oc}}$ for $i = 0,1,2,\dots,Q$. The current vector $\mathbf{i}_n$ can be obtained using 
\begin{equation}
	\label{i_n_IMPM}
	\vspace{-0.1cm}
	\mathbf{i}_n = 
	\begin{bmatrix}
		\mathbf{i}^\mathbf{E}_{n} \\ \mathbf{i}^\mathbf{I}_{n}
	\end{bmatrix}
	= \frac{1}{\sqrt{Z_\mathrm{in}(\mathbf{x},\mathcal{S})}}
	\begin{bmatrix}
		1 \\ -\left[ \mathbf{Z_I} + \mathbf{Z^L}(\mathbf{x},\mathcal{S}) \right]^{-1} \mathbf{Z_{IE}}
	\end{bmatrix}.
\end{equation}
Thus the covariance matrix $\bm{\varrho} \in \mathbb{C}^{N\times N}$ of $N$ states can be calculated through (\ref{Pattern_Correlation}), with the $(i,j)$-th entry $[\bm{\varrho}]_{i,j} = \rho_{i,j}$.

With the proposed geometry and circuit model, equations (\ref{Zin})  and (\ref{total_Pattern}) can be used to obtain the impedance and radiation pattern of the PRA-FAS for a given  $\mathbf{x}$ and set ${\cal S} $. The next step is to find the positions of the switches ${\cal S} $, and the states of the $\mathbf{x}$ that best meet the PRA-FAS design objective defined at the end of Section \Rmnum{2}. This is described in the next section.

\section{PRA-FAS Analysis and Design}

Using the expressions for impedance and radiation patterns from the previous section, we need to analyze the performance of the antenna for all possible states $\mathbf{x}$ and switch positions (set ${\cal S} $) to obtain those meeting our PRA-FAS design objective. However, the number of possible states $\mathbf{x}$ and switch sets ${\cal S} $ is very large. Taking the PRA-FAS model in Fig. \ref{Antenna_3D} as an example, there are $2^{60}$ possible states $\mathbf{x}$ with $Q=60$ internal ports. Furthermore, there are $C_{6}^{60}$ possible sets $\mathcal{S}$ or locations for the switches if there are $P=6$ switches. The total number of possible configurations that need to be analyzed is therefore $2^{60} C_{6}^{60}$, i.e., there are a huge number of possible configurations. As a result, it is necessary to develop a method that can efficiently select the appropriate internal port states $\mathbf{x}$ and sets ${\cal S}$.

In this work, we perform this by a two-step process. In the first step, we find a subset of states $\mathbf{x}$ and sets ${\cal S} $ that provide the necessary impedance match over the specified bandwidth. In the second step, we then search through these to find 12 switch states to find their optimum selection and order, that can meet our design objective. Taking the PRA-FAS design in Fig. \ref{Antenna_3D} and Fig. \ref{Pixel_layer}, where $P=6, Q=60$, these two steps are described next. 

\subsection{Step 1: Selection of Matched States}

For a given $\mathbf{x}_l$ and $\mathcal{S}_k$, the pixel surface is completely defined. However, the switches at locations $\mathcal{S}_k$ can be reconfigured to $2^6$ possible states, and therefore we can reconfigure $\mathbf{x}_l$ to any one of these $2^6$ states easily. With this in mind, we find it convenient to define a set $\mathcal{U}_{k,l}$ containing all the $2^6$ states for a given $\mathbf{x}_l$ and $\mathcal{S}_k$. That is, the set $\mathcal{U}_{k,l} $ contains those states $\mathbf{x}_l$ where the elements at positions $\mathcal{S}_k$ can take all possible values. There are $C_{6}^{60}$ sets $ \mathcal{U}_{k,l} $ for a given $\mathbf{x}_l$, and in total there are $2^{54} C_{6}^{60}$ possible sets $ \mathcal{U}_{k,l} $ across all possible states.

Not all $2^6$ states in set $ \mathcal{U}_{k,l} $ will have the required impedance matching across the desired bandwidth. This is because the driven element will adversely affect matching for some pixel combinations. Therefore, in this step, our task is to select only those sets $ \mathcal{U}_{k,l} $ with sufficient matched states. For $ \mathcal{U}_{k,l} $ to be considered a desirable set, the number of matched states in the set should exceed $N$. With this consideration, we define $\mathcal{U}^\mathrm{M}_{k,l}$ as the subset of states in set $ \mathcal{U}_{k,l} $ that are matched. This can be defined as
\begin{equation}
	\label{set_Um}
	\mathcal{U}^\mathrm{M}_{k,l} = \{ \mathbf{x}_l^m \ | \ S_\mathrm{E}(\mathbf{x}_l^m,\mathcal{S}_k) < -10 \ \mathrm{dB},\ \mathbf{x}_l^m \in  \mathcal{U}_{k,l}  \},
\end{equation}
such that 
\begin{equation}
	\label{GA1_expression}
	(C1): \ \mathrm{card} (\mathcal{U}^\mathrm{M}_{k,l}) =M \geq N,
\end{equation}
where $S_\mathrm{E}(\mathbf{x}_l^m,\mathcal{S}_k)$ is the reflection coefficient of a state
\begin{equation}
	\label{S11}
	S_\mathrm{E}(\mathbf{x}_l^m,\mathcal{S}_k) = 20 \cdot \lg \left| \frac{Z_\mathrm{in}(\mathbf{x}_l^m,\mathcal{S}_k)-\mathrm{Z}_0}{Z_\mathrm{in}(\mathbf{x}_l^m,\mathcal{S}_k)+ \mathrm{Z}_0} \right| \ \mathrm{dB},
\end{equation}
in which $\mathrm{Z}_0$ is the characteristic impedance. 

The set  $ \mathcal{U}^{\mathrm{M}}_{k,l} $ is then formally defined as those that meet (\ref{set_Um}) and constraint (\ref{GA1_expression}) in the remainder of this paper.

The number of sets  $ \mathcal{U}^{\mathrm{M}}_{k,l} $ could be quite large. However it is not necessary to find all sets $ \mathcal{U}^{\mathrm{M}}_{k,l} $ that meet the matching requirement. 
Only enough sets are required that provide a sufficiently high chance that the selection and ordering objective in the second step can be met. 

Typically, we have found that we only need to obtain approximately 100 matched sets $ \mathcal{U}^{\mathrm{M}}_{k,l} $  to proceed to the next step. Therefore, step 1 serves the purpose of not only finding those configurations that are matched, but it is also used to reduce the search space for the next step, from $2^{54} C_{6}^{60}$ to 100.


\subsection{Step 2: Meeting the Spatial Covariance Requirement}

From the matched sets $ \mathcal{U}^{\mathrm{M}}_{k,l} $ gained in step 1, we must now select the best set that meets the spatial correlation objective (\ref{Covariance_ij}) for $N$ out of its $M$ matched states in $ \mathcal{U}^{\mathrm{M}}_{k,l} $. Furthermore, this step also requires an ordering process where the $N$ states are mapped to the FAS port number that best meets (\ref{Covariance_ij}).    

To start step 2, the correlations between the states in each set $\mathcal{U}^{\mathrm{M}}_{k,l} $ need to be found. To perform this, all $M$ radiation patterns must be obtained through (\ref{total_Pattern}) for each set $ \mathcal{U}^{\mathrm{M}}_{k,l}$. This can impose a heavy burden on computation, since it will also require (\ref{Pattern_Correlation}) to be calculated $M(M-1)/2$ times for each $ \mathcal{U}^{\mathrm{M}}_{k,l}$. To make this process more computationally efficient, we can leverage previous work on a related pattern correlation decomposition method (PCDM) \cite{PCDM}.



In PCDM, all current vectors of the $M$ reconfigurable states can be written compactly by collecting them into a matrix as $\mathbf{I}=\left[\mathbf{i}_1,\mathbf{i}_2,\dots,\mathbf{i}_M\right]$, and all total radiation patterns can similarly be compactly written as $\mathbf{E}=\left[\mathbf{e}_1,\mathbf{e}_2,\dots,\mathbf{e}_M \right]$. The relationship between $\mathbf{E}$ and $\mathbf{I}$ is then given by
\begin{equation}
	\label{total_oc}
	\mathbf{E}=\mathbf{E}_{\mathrm{oc}} \mathbf{I}.
\end{equation}
This allows the pattern covariance matrix $\bm{\varrho}_0$ of all $M$ PRA-FAS states to be written as
\begin{equation}
	\label{Rho_Matrix}
	\boldsymbol{\varrho}_0 
	=\left|\mathbf{C} \oslash \mathbf{G}\right|,
\end{equation}
where $\oslash$ is Hadamard division, and $\mathbf{C}$ is the correlation matrix of all the $M$ patterns, defined as
\begin{equation}
	\label{C_expression}
	\begin{split}
		\mathbf{C}
		&=\iint_\mathbf{\Omega} \mathbf{E}^\mathrm{H} \mathbf{E}  S(\mathbf{\Omega}) \mathrm{d} \mathbf{\Omega}
		=\iint_\mathbf{\Omega} \mathbf{I}_{}^\mathrm{H} \mathbf{E}_{\mathrm{oc}}^\mathrm{H} \mathbf{E}_{\mathrm{oc}} S(\mathbf{\Omega}) \mathbf{I} \mathrm{d} \mathbf{\Omega}  \\
		&=\mathbf{I}_{}^\mathrm{H} \left(\iint_\mathbf{\Omega} \mathbf{E}_{\mathrm{oc}}^\mathrm{H}  \mathbf{E}_{\mathrm{oc}}  S(\mathbf{\Omega}) \mathrm{d} \mathbf{\Omega} \right) \mathbf{I}
		=\mathbf{I}_{}^\mathrm{H} \mathbf{K_\mathrm{oc}} \mathbf{I},
	\end{split}
\end{equation}
where $\mathbf{K}_\mathrm{oc}\in \mathbb{C}^{(Q+1)\times (Q+1)}$ is the correlation matrix of all open-circuit radiation patterns weighted by the PAS. The matrix $\mathbf{G}\in \mathbb{C}^{M\times M}$ represents the average energy of the patterns, which is used for normalization. The $(i,j)$-th entry of $\mathbf{G}$ is written as
\begin{equation}
	\label{G_element}
	\left[\mathbf{G}\right]_{i,j}=\sqrt{\left[\mathbf{C}\right]_{i,i}\left[\mathbf{C}\right]_{j,j}}.
\end{equation}

The expression (\ref{C_expression}) provides a direct method to find the pattern correlations, while (\ref{total_Pattern}) is no longer required. This is because $\mathbf{K}_\mathrm{oc}$ only needs to be found once for the PRA-FAS. Then all the spatial correlations for all searched pixel configurations can be found using (\ref{Rho_Matrix}) to (\ref{G_element}) straightforwardly, because the currents $\mathbf{I}$ have already been found in determining the PRA-FAS impedance. 

The next task in step 2 is the selection of $N=12$ states from the $M$ available in each $\mathcal{U}^\mathrm{M}_{k,l}$ so that the spatial correlation objective (\ref{Covariance_ij}) is best met. This task is made intricate because both the selection and ordering of the states are important in meeting the spatial correlation objective (\ref{Covariance_ij}). As such, a vector sequence $\mathbf{D} = [d_1, d_2,\dots,d_N ]^\mathrm{T}$ is used to capture the order and selection of $N$ states from $M$. Each element in $\mathbf{D}$ is in the range $1$ to $M$, and each value can only be used once. There are $N!C_N^M$ combinations for the sequence of states for $\mathbf{D}$. The $n$-th element of $\mathbf{D}$ acts as a map from the $n$-th FAS port to the PRA-FAS state $[\mathbf{D}]_n$. We wish to search through all possible $\mathbf{D}$ to find the sequence that produces the closest match to (\ref{Covariance_ij}).

Quantifying the match can be performed by defining a function as the difference between the covariance matrix generated by our PRA-FAS design and the target covariance matrix. The target covariance matrix used here is defined as $\bm{\varrho^*}$, with its $(n,n')$-th entry given by
\begin{equation}
	\label{target_covariance_matrix}
	\left[ \bm{\varrho^*} \right]_{n,n'} = \rho_{n,n'} = J_0 \left( \frac{2 \pi |n-n'|W}{N-1} \right) .
\end{equation}
According to the given $\mathbf{D}$, we take the terms in $\bm{\varrho}_0$ in (\ref{Rho_Matrix}), to form the covariance matrix $\bm{\varrho} \in \mathbb{C}^{N \times N}$ of $N$ states. Since we are not interested in the phase of the FAS port correlation, we use absolute value for a particular order $\mathbf{D}$ as
\begin{equation}
	\label{GA2_expression_obj}
	\Delta(\mathbf{D}) = \sum_{n=1}^N \sum_{n'=1}^N \left|  \left|\left[\bm{\varrho(D)}\right]_{n,n'} \right| - \left| \left[\bm{\varrho^*}\right]_{n,n'}\right| \right|.
\end{equation}



It is also important to explicitly include frequency so that design specifications for bandwidth can also be met. Assuming the target specification is a single band with the lower limit, $f_l$, and the upper limit, $f_u$, we can evaluate (\ref{GA2_expression_obj}) at $T$ frequency samples, with $f_t= f_l + (t-1)\frac{f_u-f_l}{T-1}$ for $t=1,2,\dots,T$. Explicitly including frequency dependence in our expressions, results in (\ref{set_Um}) becoming
\begin{equation}
	\label{set_Um_update}
	\mathcal{U}^\mathrm{M}_{k,l} = \{ \mathbf{x}_l^m \ | \ \max \left[ S_\mathrm{E}(\mathbf{x}_l^m,\mathcal{S}_k,f_t) \right] < -10 \mathrm{dB},\ \mathbf{x}_l^m \in \mathcal{U}_{k,l} \}
\end{equation}
for $t=1,2,\dots,T$. Correspondingly, the total absolute error (\ref{GA2_expression_obj}) also becomes
\begin{equation}
	\label{GA2_updated}
	\ \Delta(\mathbf{D}) =  \sum_{t=1}^T \sum_{n=1}^N \sum_{n'=1}^N \left|  \left|\left[\bm{\varrho}(\mathbf{D},f_t)\right]_{n,n'}\right| - \left| \left[\bm{\varrho^*}\right]_{n,n'}\right| \right|.
\end{equation}

To quantify the proximity of our design results to the ideal target $\bm{\varrho^*}$, the average error $\delta_e$ can be selected as the objective function. This is defined as the total absolute error divided by the number of entries in the covariance matrix, taking into account the $T$ frequency samples. It is a measure of the average difference of the covariance of an element from the desired value and is always less than unity and should be as small as possible and at least less than 0.10. The resulting objective function is given by
\begin{equation}
	\label{relative error}
	\delta_e(\mathbf{D}) = \frac{\Delta(\mathbf{D})}{T N^2}.
\end{equation}

\begin{algorithm}[!t]
	\caption{Decode Algorithm for Non-repeating Order}
	\begin{algorithmic}
		\STATE 
		\STATE {\textbf{Input:}} \ $\mathbf{B}=\left[ b_1,b_2,\dots,b_N\right] \in \{ 1,2,\dots,M \}^{N}$
		\STATE \hspace{0.5cm}$ \textbf{Initialization:} \ i = 0, \ \mathbf{H}=\left[ 1,2,\dots,M \right]$
		\STATE \hspace{2.5cm}$\mathbf{D}=\left[ d_1,d_2,\dots,d_N\right]=\textbf{0}$
		\STATE \hspace{0.5cm}$ \textbf{Define:}\ \dim(\cdot) \ \text{expresses the length of the vector.}$
		\STATE \hspace{0.5cm}$ \textbf{repeat}$
		\STATE \hspace{0.5cm}$ i = i+1$
		\STATE \hspace{1cm}$ \text{Calculate} \ j=\left[ (b_i-1) \mod \dim(\mathbf{H}) \right] +1$
		\STATE \hspace{1cm}$ \text{Update} \ d_i =\left[ \mathbf{H}\right]_j $
		\STATE \hspace{1cm}$ \text{Delete} \ \left[ \mathbf{H}\right]_j \ \text{from the vector} \ \mathbf{H}$
		\STATE \hspace{0.5cm}$ \textbf{until} \ i \geq N$
		\STATE {\textbf{Output:}} \ $\mathbf{D}=\left[ d_1,d_2,\dots,d_N\right],d_n \in \{ 1,2,\dots,M \},$
		\STATE \hspace{1.5cm} $ d_n \neq d_{n'} \ \text{for any} \ n \neq n'.  $
	\end{algorithmic}
	\label{Decode_step}
\end{algorithm}

\begin{table}[!t]
	\caption{Parameters Used in GA}
	\label{GA2} 
	\centering
	\begin{tabular}{cc}
		\toprule
		Parameter & Value \\
		\midrule
		Max. Generations & 200 \\
		Population Size & 600 \\
		Cross Probability & 0.8 (Uniformly Distributed) \\
		Mutation Probability & 0.1 (Uniformly Distributed) \\
		\bottomrule
	\end{tabular}
\end{table}

Our optimization problem can then be written as 
\begin{equation}
	\label{GA2_expression}
	\begin{split}
		& \mathop{\min} \limits_{\mathbf{D}} \ \delta_e(\mathbf{D})\\
		\mathbf{s.t.} \ (C2): \ &\mathbf{D} \in \{1,2,\dots,M\}^N, \mathrm{with} \ \left[ \mathbf{D} \right]_n \neq \left[ \mathbf{D} \right]_{n'}.
	\end{split}
\end{equation}
This combinatorial optimization (\ref{GA2_expression}) is an NP-hard problem. Therefore, a heuristic algorithm, GA \cite{IMPM,GA_try1,GA_try2} is adopted here to solve it. Detailed parameters in the use of GA are given in Table \ref{GA2}, where our approach provides computational efficiency, allowing for rapid optimization.

Using GA, the optimum switch locations $\mathcal{S}^*_k$, fixed hardwire configuration $\mathbf{x}^{*}_l$ (excluding those elements in $\mathcal{S}^*_k$ which are switches), as well as the mapping $\mathbf{D}^*$ are all found. These can then be used to find the design of the required PRA-FAS. 

When using GA to optimize the selection and ordering of states, it should be noted that we need to be careful about the crossovers between individual chromosomes (variables). This is because duplicated elements within the sequence, $\mathbf{D}$ cannot directly serve as the new chromosome. Here we propose an encoding method that can transform a coded vector $\mathbf{B}$, which may contain repeated elements, into the corresponding non-repeating vector $\mathbf{D}$. The steps of the decoding process are depicted in Algorithm \ref{Decode_step}, ensuring the resulting sequence $\mathbf{D}$ remains valid and free from duplication. 

\begin{figure}[!t]
	\centering
	\includegraphics[width=1\linewidth]{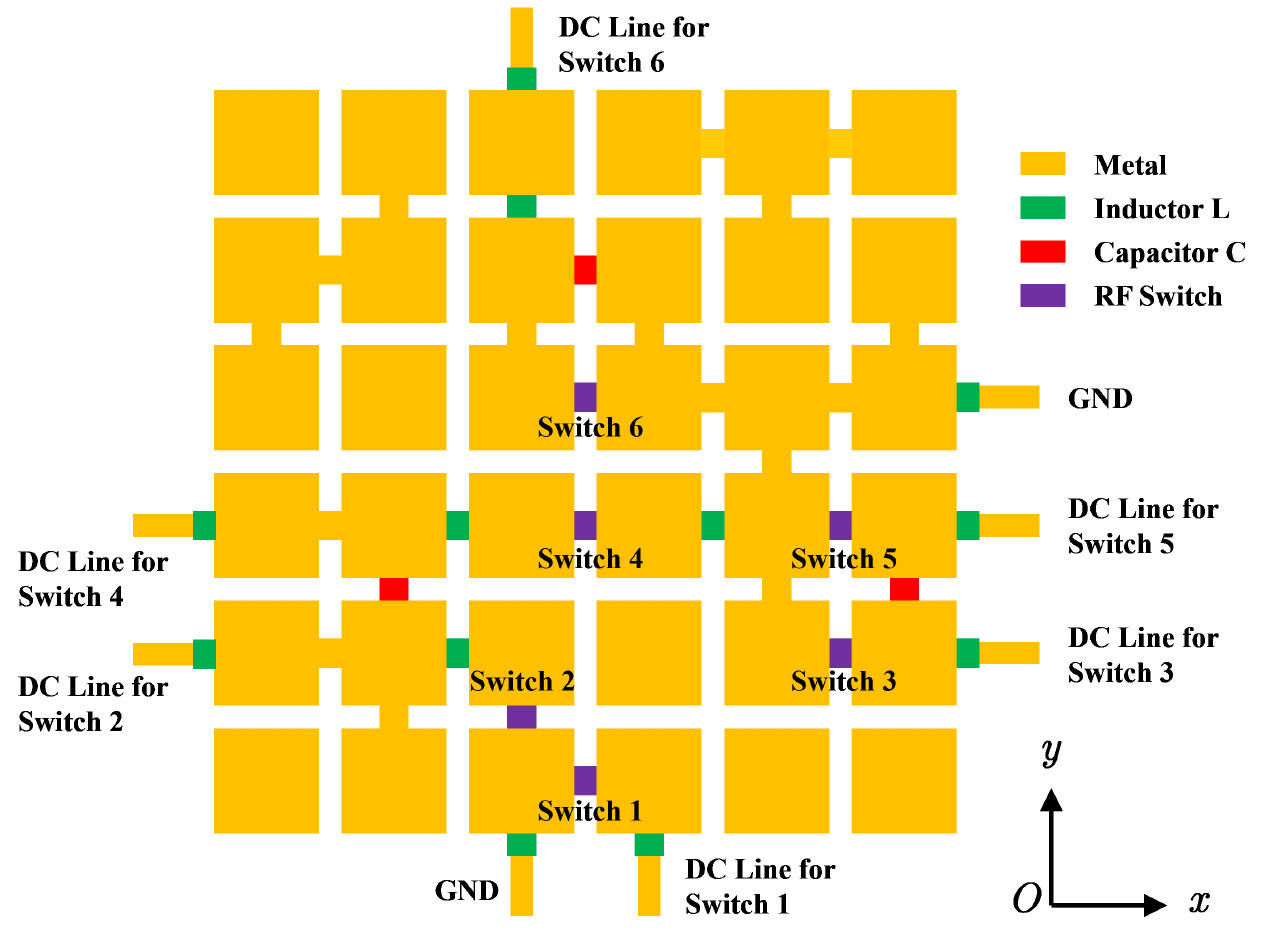}
	\caption{Optimized configuration of the PRA-FAS, consisting of the positions of 6 switches and shorted internal ports (connected by hardwires). Further details about the design and impact of the DC lines are provided in the text. Unconnected pixels in the corners cannot be removed, since they form $\mathbf{Z}$ together with other pixels and they are also part of the optimization process.}
	\label{Switch_Position}
\end{figure}

\section{Simulation and Measurement Results}

In this section, simulation and measurement results for our PRA-FAS design operating at the center frequency of 2.5 GHz are provided. To ensure the FAS-PRA has uni-directional radiation, the PAS was selected as $S\mathbf{(\Omega)}$ with $S\mathbf{(\theta,\phi)}=S_0$ for $\theta \in [0,\pi/2], \phi \in [0,2\pi)$ and $S\mathbf{(\theta,\phi)}=0$ otherwise in all simulations.

In all the simulations, CST Studio Suite \cite{cstStudioSuite} is used to obtain full-wave simulations of S-parameters, radiation patterns, and efficiency. 

We also need to specify the design of the switches so that the simulations can include their effects and the PRA-FAS can be fabricated in practice. In the design examples, we utilize PIN diodes for the switches similar to other approaches to PRA design \cite{pixel0,pixel1,pixel2,Jerry1,Jerry2}. We use PIN diode MA4ACP907 \cite{datasheet907} and its equivalent circuit models for on and off are shown in Fig. \ref{Circuit_of_switch}. The control of the PIN diodes is performed using DC signals, and therefore isolation between the DC signals and RF signals is required. This can be provided by using auxiliary inductors and capacitors. Inductors 0402DC-R10XJRW \cite{coilcraft} from Coilcraft serve as RF chokes, equivalent to RF open-circuit impedance connected to the internal ports. Capacitors GJM1555C1H130GB01 \cite{murata} from Murata are utilized as both RF short for internal ports and block for DC signals. In order to control the PIN diodes without interfering with the operation of PRA-FAS, the DC control feeds are all located around the boundary of the PRA-FAS. Considering that the DC feed lines are inductively isolated from the radiating structure (pixels), and their length is much smaller than the wavelength, the DC feed minimally affects the RF currents as well as radiation. The switch layout and the corresponding auxiliary components for our PRA-FAS are shown in Fig. \ref{Switch_Position}. This design and the simulation will be discussed further in the next section. 

\begin{table}[!t]
	\caption{Optimized State Order of the 6 Switches}
	\label{Switch_states}
	\centering
	\begin{tabular}{ccccccc}
		\toprule
		Switch No. (in Fig. \ref{Switch_Position}) & 1 & 2 & 3 & 4 & 5 & 6\\
		\midrule
		Switch Position ($\mathcal{S}^*_k$) & 3 & 8 & 16 & 25 & 27 & 36 \\ 
		\midrule
		State 1 & on & on & off & off & off & off\\
		State 2 & on & on & off & off & on & off\\
		State 3 & on & on & on & off & on & off\\
		State 4 & off & off & off & off & on & off\\
		State 5 & off & on & on & off & on & off\\
		State 6 & off & on & off & on & on & off\\
		State 7 & off & on & on & on & on & off\\
		State 8 & off & off & on & on & on & on\\
		State 9 & off & on & on & on & on & on\\
		State 10 & off & on & on & on & off & on\\
		State 11 & off & on & on & off & off & on\\
		State 12 & off & off & on & off & off & on\\
		\bottomrule
	\end{tabular}
\end{table}

\begin{figure}[!t]
	\centering
	\includegraphics[width=1\linewidth]{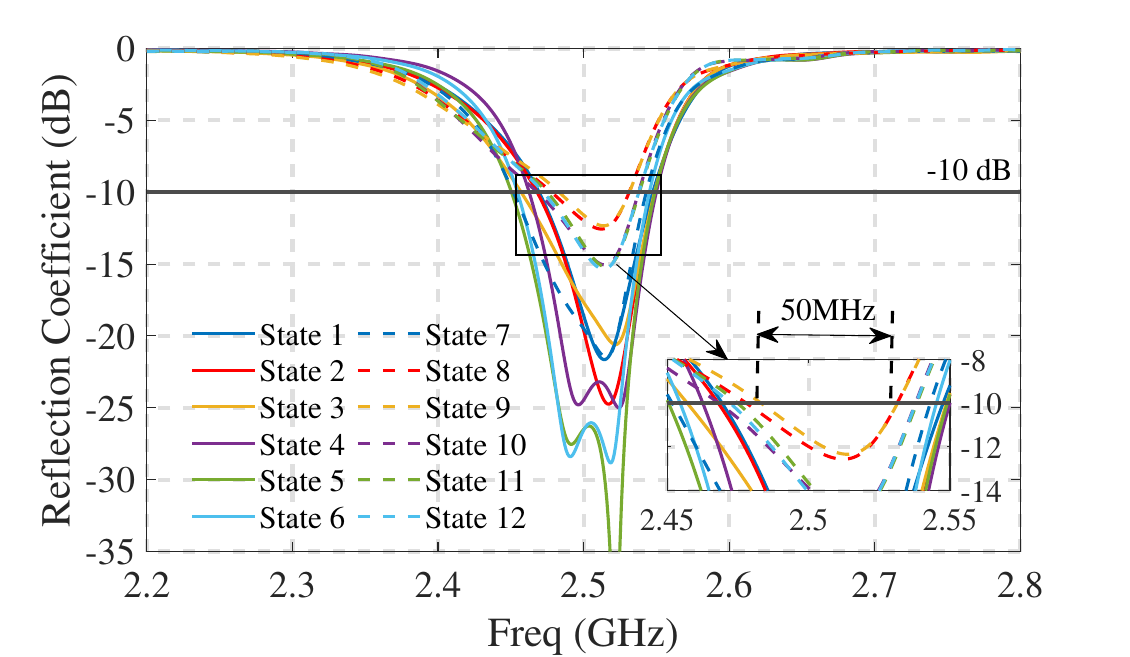}
	\caption{Simulated reflection coefficients of 12 states of the proposed PRA-FAS in Fig. \ref{Antenna_3D} versus frequency.}
	\label{S11_sim_12}
\end{figure}

\begin{figure*}[!t]
	\centering
	\subfloat[]{\includegraphics[width=0.32\linewidth]{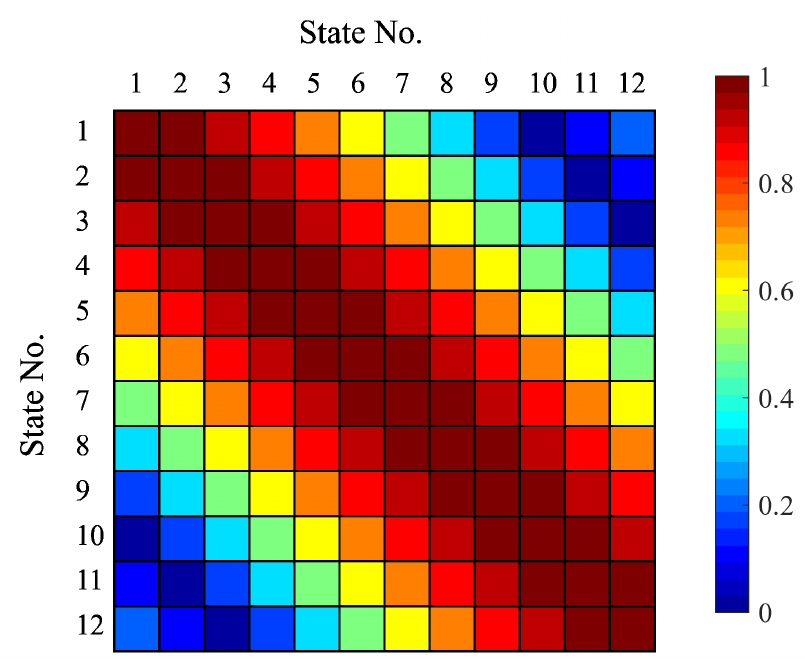}%
		\label{CorrelationMatrix_Ideal}}
	\hfil
	\subfloat[]{\includegraphics[width=0.32\linewidth]{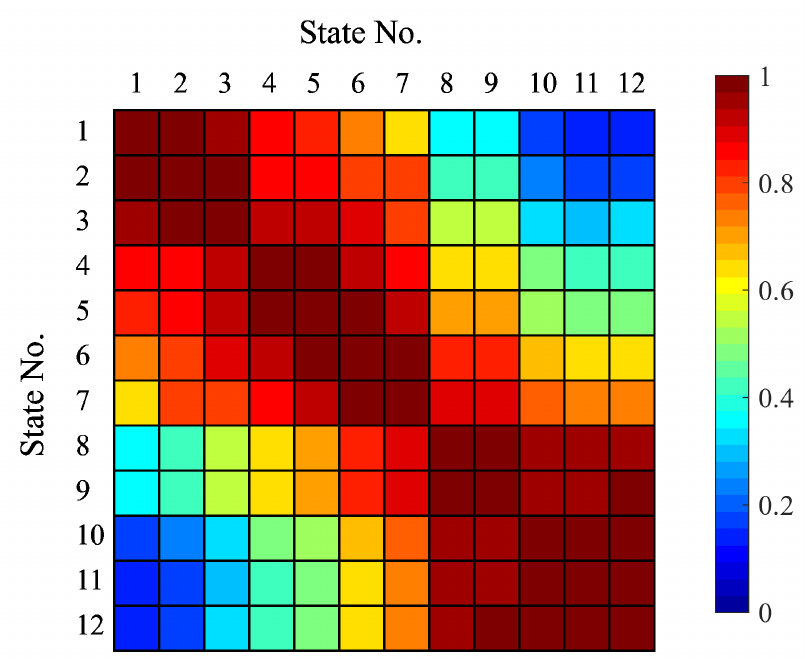}%
		\label{CorrelationMatrix_Sim}}
	\hfil
	\subfloat[]{\includegraphics[width=0.32\linewidth]{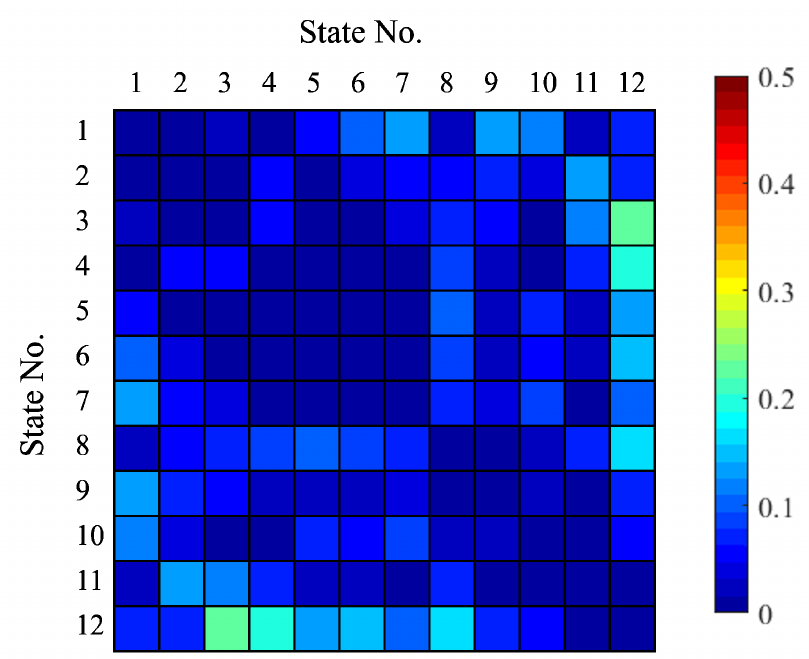}%
		\label{CorrelationMatrix_Error}}
	\caption{Covariance matrices for $N=12$ states of the proposed PRA-FAS with $W=0.5$. (a) The ideal target covariance matrix. (b) The simulated covariance matrix using our proposed design, with an average relative error of $\delta_e=0.063$. (c) The absolute error between the simulated result and the target covariance matrix.}
	\label{CorrelationMatrix}
\end{figure*}

\subsection{Simulation Results}

In the first simulation, we provide results for our reference PRA-FAS design as given in Fig. \ref{Antenna_3D}, where we target operation at the center frequency of 2.5 GHz. We set $T=1$ to illuminate the basic design process without the complication of bandwidth optimization. Following the PRA-FAS design process described in the previous sections, we obtain a PRA-FAS design where the optimum switch locations $\mathcal{S}^*_k$, fixed hardwire configuration $\mathbf{x}^{*}_l$ (excluding those elements at $\mathcal{S}^*_k$ which are switches), as well as the corresponding mapping $\mathbf{D}^*$ are all found.

The resulting design is shown in Fig. \ref{Switch_Position}, where the detailed structure of the DC control lines, RF chokes, and capacitors that we have devised can be seen. The states of each switch corresponding to the 12 states of the PRA-FAS arranged in order is shown in Table \ref{Switch_states}.

First, we provide the simulated S-parameters in Fig. \ref{S11_sim_12} to verify that the antenna is operational in all its states. From these, we can see that the PRA-FAS is well-matched to the center frequency of 2.5 GHz, and the overall bandwidth is slightly over 50 MHz.

The key design parameter for FAS is however the correlation characteristics of the design. In Fig. \ref{CorrelationMatrix}(a), we first show the target covariance objective $\bm{\varrho^*}$ as specified by
(\ref{target_covariance_matrix}) when $N=12$ and $W=0.5$. This figure presents the covariance matrix for $n,n' \in \{1,2,\dots,12\}$. As expected, the leading diagonal is unity and the matrix is symmetric. Fig. \ref{CorrelationMatrix}(b) presents the simulated pattern covariance matrix of the optimized 12 states of the PRA-FAS obtained by simulation. These are obtained from simulations of the open circuit patterns and using (\ref{target_covariance_matrix}). Fig. \ref{CorrelationMatrix}(c) presents the difference between the simulated covariance matrix and the target correlation. It can be seen that the covariance values between any pair of patterns are found to be very near to the intended target. Using the operating states in Table \ref{Switch_states}, the minimum average relative error between the simulated covariance matrix and target covariance is as low as $\delta_e =  0.063$ (see equation (\ref{relative error})), thereby demonstrating the precision of the proposed PRA-FAS design method.

\begin{figure}[!t]
	\centering
	\includegraphics[width=1\linewidth]{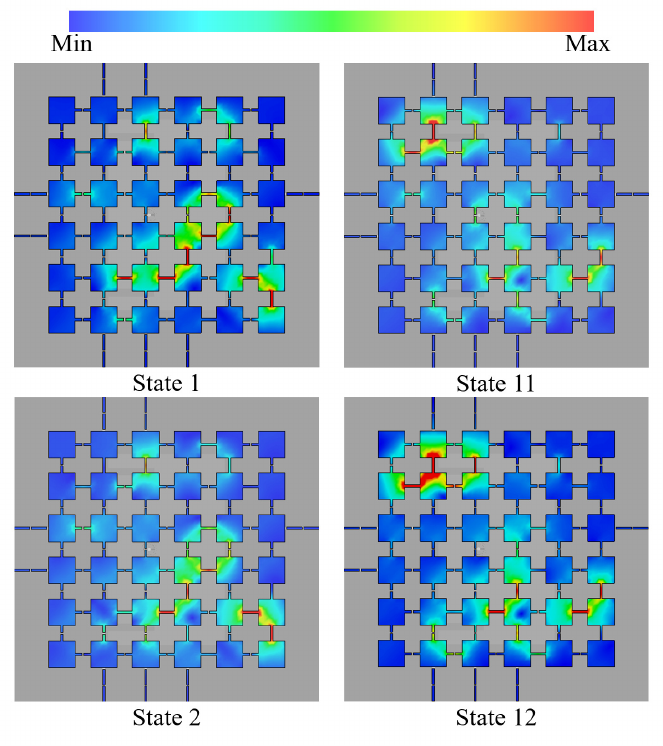}
	\caption{Simulated pixel layer current distribution of PRA-FAS states numbered 1,2,11,12, respectively, at 2.5 GHz.}
	\label{Pattern_4}
\end{figure}


From the covariance matrix of the PRA-FAS, it can be seen that adjacent FAS states (ports) have a high correlation, which implies that the radiation patterns of these adjacent states should exhibit a greater degree of similarity based on formula (\ref{Pattern_Correlation}). To substantiate this, we refer to the simulation results of the pixel layer current distribution for 4 FAS states. In Fig. \ref{Pattern_4}, the current distribution on states 1, 2, 11, and 12 are shown. It is evident that adjacent states, such as 1 and 2, or 11 and 12, share similar current distributions. Consequently, their radiation patterns are indeed also similar. In contrast, non-adjacent states, like 1 and 11, or 1 and 12, exhibit significant differences in current distributions, leading to uncorrelated radiation patterns, which aligns with our theoretical predictions. 

\begin{figure}[!t]
	\centering
	\subfloat[]{\includegraphics[width=0.5\linewidth]{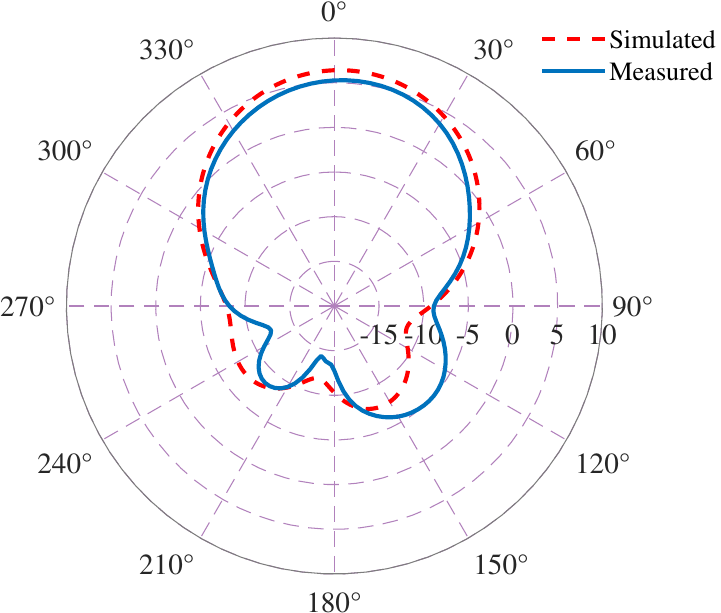}}
	\hfil
	\subfloat[]{\includegraphics[width=0.5\linewidth]{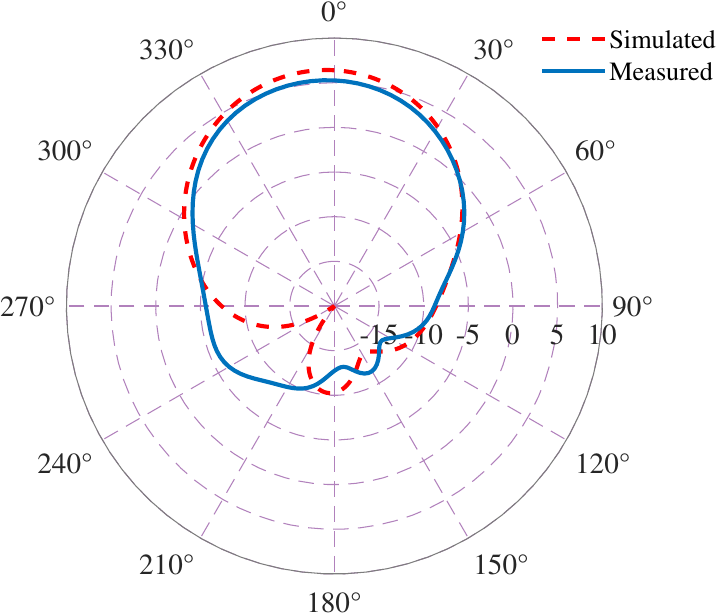}}
	\caption{Simulated and measured radiation patterns of State 1 at 2.5 GHz with (a) $\phi = 0^\circ$ (b) $\phi = 90^\circ$ (Unit: dBi).}
	\label{Radiation_State1}
\end{figure}

We can also find from Fig. \ref{Pattern_4} that the current distribution is not symmetrical. This asymmetry implies that the radiation pattern is also not symmetrical, causing the main beam direction to deviate from the central normal of the radiation aperture. Taking the simulated radiation pattern of state 1, given in Fig. \ref{Radiation_State1}, as an example, its maximum gain (7.14 dBi) is not located on the symmetry plane of $\phi=0^\circ$ and $\phi=90^\circ$. Similar deviations are observed for the remaining 11 states. However, for all 12 states, the angle of the peak deviation from the center normal does not exceed $15^{\circ}$. Despite these deviations, it remains accurate to characterize all 12 states as approximately uni-directional radiating.

It should be noted that in finding the signal correlation using (\ref{Pattern_Correlation}), it is the difference between the radiation patterns that reduce the correlation. Even though the magnitudes of the simulated and measured radiation patterns of all states are similar, it is mainly the difference between the phases that result in the various correlations between states in this antenna. An example of this effect can be seen for the dipoles in (\ref{eqn-dipole}), where only the pattern phase is changed.



\begin{figure}[!t]
	\centering
	\includegraphics[width=0.95\linewidth]{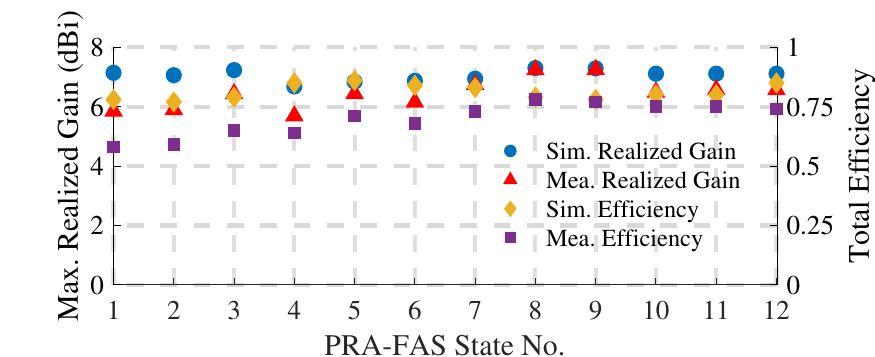}%
	\caption{Simulated, measured maximum realized gain and total efficiency of all 12 states at 2.5 GHz.}
	\label{Gain_state}
\end{figure}

Fig. \ref{Gain_state} shows the realized gains as well as the efficiencies of all 12 PRA-FAS states at the center frequency of 2.5 GHz. It can be seen that the realized gains of all states are greater than 6.6 dBi. Meanwhile, an average efficiency of around 80\% is achieved, which is sufficient for practical applications given the presence of the RF switches. 

\subsection{Experimental Results}

Experimental results are required to verify our proposed design and the proposed design process. We provide experimental results for the PRA-FAS design in Fig. \ref{Antenna_3D}. 

\begin{figure}[!t]
	\begin{minipage}[]{0.54\linewidth}
		\centering
		\subfloat[]{\includegraphics[width=0.75\linewidth]{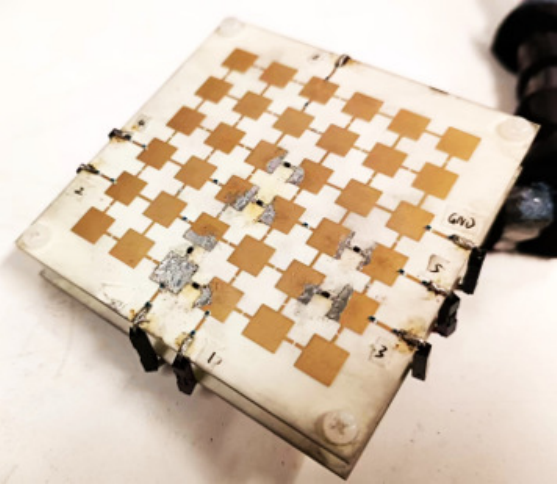}}
		\hfil
		\centering
		\vspace{-0.05cm}
		\subfloat[]{\includegraphics[width=1\linewidth]{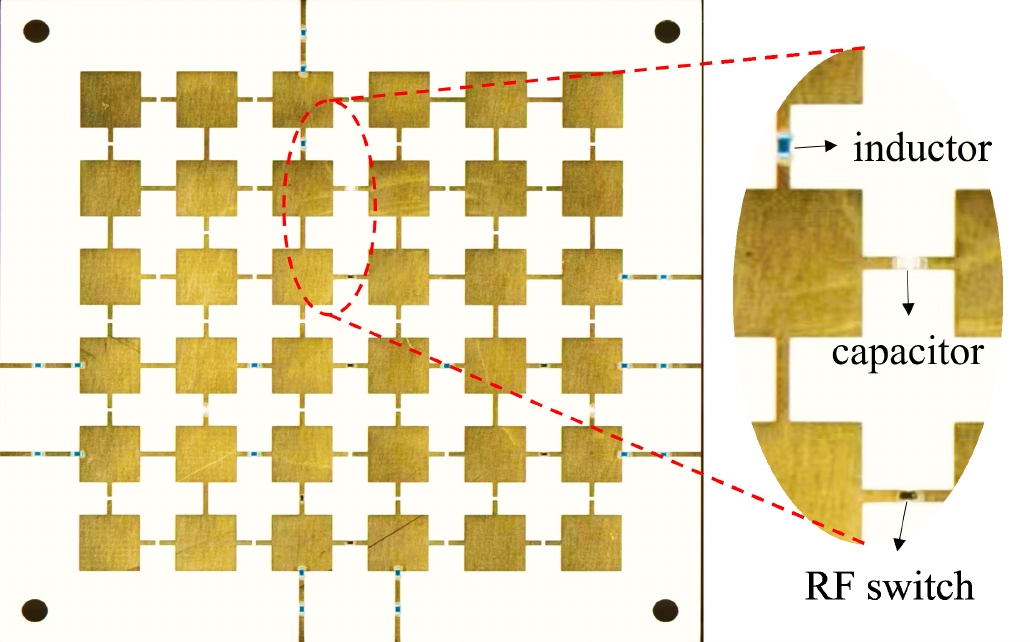}}
		\hfil
	\end{minipage}
	\begin{minipage}[]{0.45\linewidth}
		\centering
		\subfloat[]{\includegraphics[width=1\linewidth]{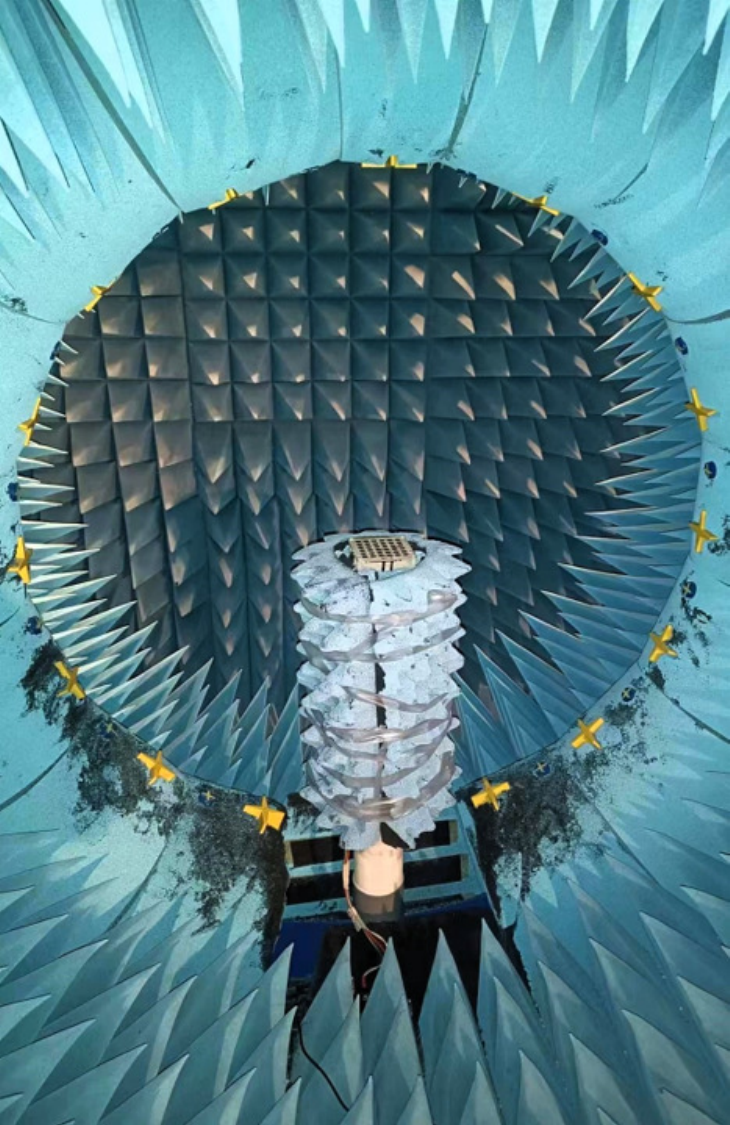}}
		\hfil
	\end{minipage}
	\caption{(a) Prototype of the proposed PRA-FAS. (b) Top view of the pixel layer. (c)  Measured setup: the proposed PRA-FAS is controlled by an FPGA.}
	\label{PRA_Real}
\end{figure}

\begin{figure}[!t]
	\centering
	\includegraphics[width=0.9\linewidth]{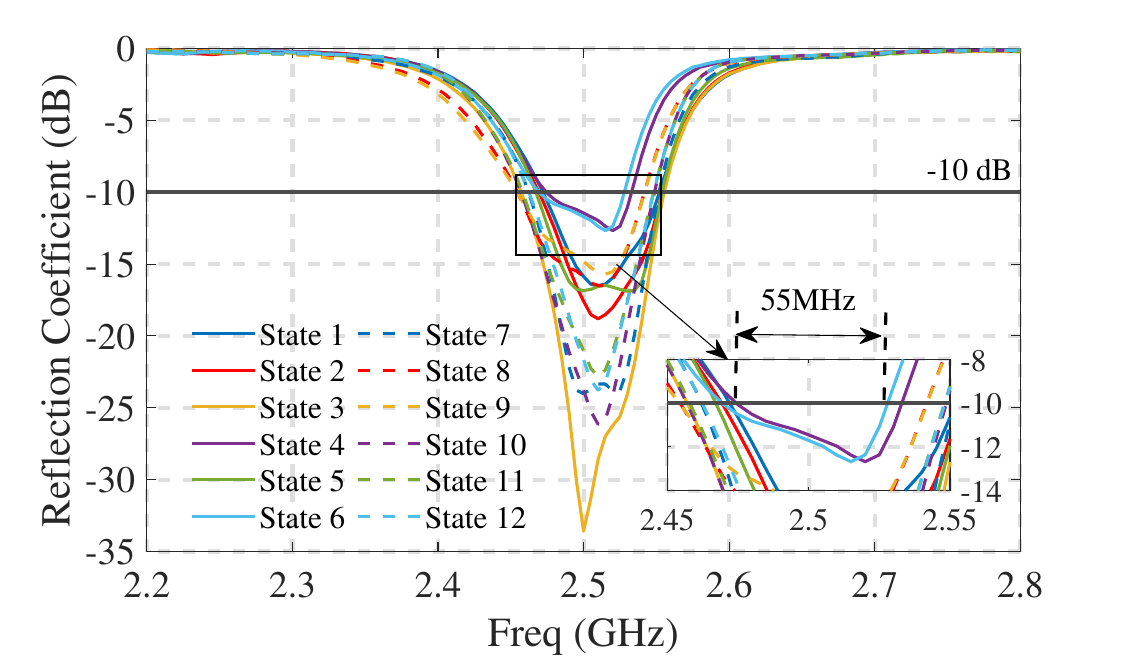}
	\caption{Measured reflection coefficients of 12 states of the PRA-FAS prototype in Fig. \ref{PRA_Real}(a) versus frequency.}
	\label{S11_mea_12}
\end{figure}

The photograph of the experimental prototype is presented in Fig. \ref{PRA_Real}(a). The pixel structure on the upper layer of the PRA-FAS is given in \ref{PRA_Real}(b). An FPGA AX7035 (0-/3.3-V output) is deployed to control the states of all 6 switches, achieving switching among 12 operation states. Fig. \ref{PRA_Real}(c) shows the test setup of the PRA-FAS, where absorbing materials shield the DC control lines, preventing interference with the radiation field. Fig. \ref{S11_mea_12} presents the measured reflection coefficients for all 12 states of the PRA-FAS, indicating that all states achieve impedance matching successfully and exhibit an operating bandwidth exceeding 50 MHz, similar to the simulation results in Fig. \ref{S11_sim_12}.

\begin{figure}[!t]
	\centering
	\vspace{-0.05cm}
	\includegraphics[width=0.67\linewidth]{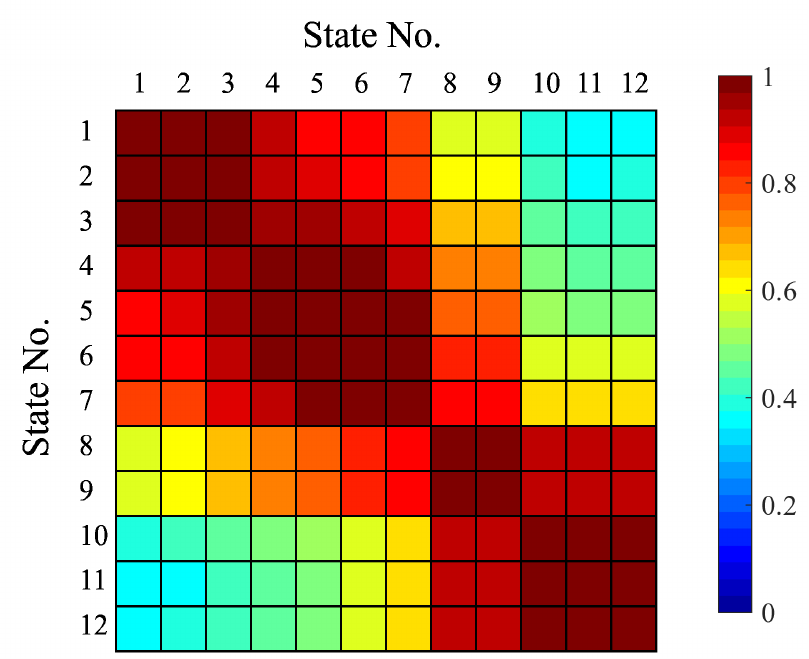}
	\caption{Measured covariance matrix of 12 PRA-FAS states. The average relative error over all correlation terms is $\delta_e=0.108$.}
	\label{CorrelationMatrix_Mea}
\end{figure}

In the correlation calculation, only the electric field components in the upper hemispherical space are adopted (see the PAS at the beginning of Section \Rmnum{5}). Fig. \ref{Radiation_State1} shows the measured radiation pattern of State 1, and its upper half is in good agreement with the simulated results. This consistency also exists in the other 11 PRA-FAS states, indicating the accuracy of the measured results. In addition, Fig. \ref{Gain_state} illustrates the measured peak gain and total efficiency of the radiation patterns across various states of the PRA-FAS. The test outcomes closely align with the simulation data, with discrepancies attributable to manufacturing variances. 

Fig. \ref{CorrelationMatrix_Mea} presents the covariance matrix obtained from measurement, with the average error of $\delta_e=0.108$ (see equation (\ref{relative error})). Compared to the simulated covariance matrix in Fig. \ref{CorrelationMatrix}(b) ($\delta_e=0.063$), there is an increase in average error in the experimental outcomes. This is most likely due to modeling errors in the PIN diodes as detailed in the discussion section next. Despite the deterioration of the average error, each column or row in the measured covariance matrix still exhibits a distinct Bessel curve trend, making it applicable to FAS scenarios.

The final litmus test for our proposed FAS is to obtain measurements of the signals at the FAS ports in a rich scattering environment. To do this, we use our $4\times4$ MIMO testbed \cite{testbed}. It can provide measurements of $4\times 4$ channels every 0.01 seconds. To utilize the test bed for FAS measurements, it was configured as a $2\times 2$ testbed. The two transmit antennas were set as dipoles and separated by over 2 wavelengths to provide uncorrelated transmit channels. At the receiver, one antenna was set as the FAS under test and the other a dipole. Both were again separated by 2 wavelengths so they were uncorrelated. In particular, receiving port 1 was set to be a dipole, and receiving port 2 was the PRA-FAS. With both ports 1 and 2 of the transmitter set as dipoles, the MIMO channel elements, $h(1,1)$ and $h(1,2)$ correspond to a regular antenna system and can be used as benchmark channels. Channels $h(2,1)$ and $h(2,2)$ correspond to the two channels with the PRA-FAS as the receiving antenna. The FAS was configured with an FPGA and set to cycle through the 12 FAS states in order. The state of the FAS could then be synchronized with the testbed so the channel for each FAS state could be identified. The MIMO antenna channel measurements were taken in the Wireless Communication Lab, at the Hong Kong University of Science and Technology. The transmitter and receiver were put in a non-line-of-sight (NLOS) configuration separated by 5 m with 2 m high cupboards blocking the line-of-sight (LOS) path. 

\begin{figure}[!t]
	\centering
	\includegraphics[width=0.9\linewidth]{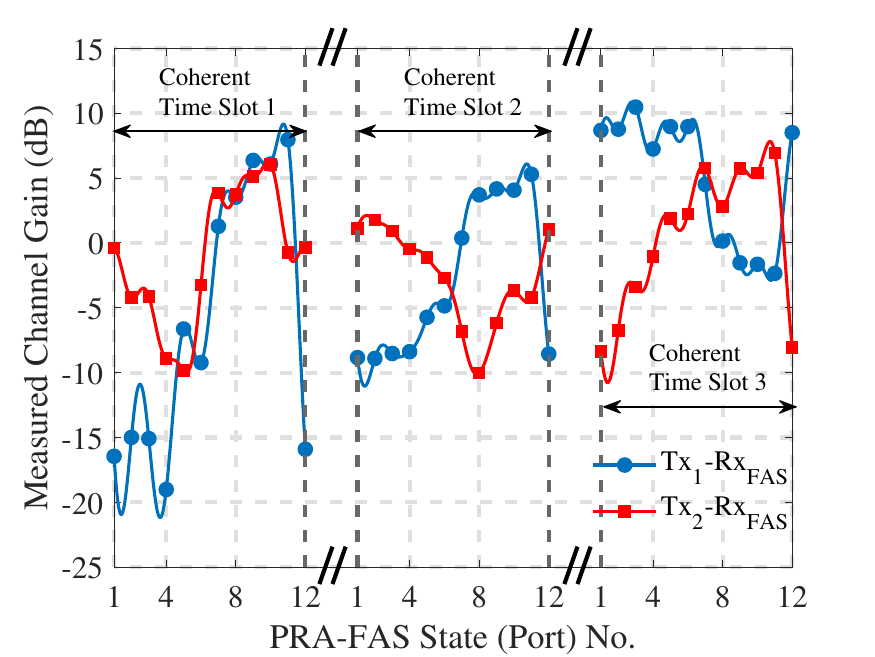}
	\caption{$h(2,1)$ and $h(2,2)$ (2 FAS channels) of 12 PRA-FAS states (ports) for three different stationary channels. The solid dots are the sample points while the interpolated lines have been added for clearer visualization.}
	\label{h_FAS}
\end{figure}

The first set of results we present are for when the rich scattering channel is stationary and the FAS cycled through its 12 configurations. The stationary of the channels could be checked by the 2 benchmark channels, $h(1,1)$ and $h(1,2)$, which did not change during the measurement interval of the 12 configuration states. Three sets of channel samples from the $h(2,1)$ and $h(2,2)$ measurements over the 12 FAS states are shown in Fig. \ref{h_FAS}. These three sets of measurements were each for a different stationary channel. It can be observed that the FAS can provide channels that vary with its configuration state as required. In particular, as we cycle through the 12 PRA-FAS states, we obtain up to around 30 dB signal variation while the benchmark channels remain stationary. This shows that the PRA-FAS is generating sufficient diversity by changing its radiation patterns.

\begin{figure}[!t]
	\centering
	\includegraphics[width=0.9\linewidth]{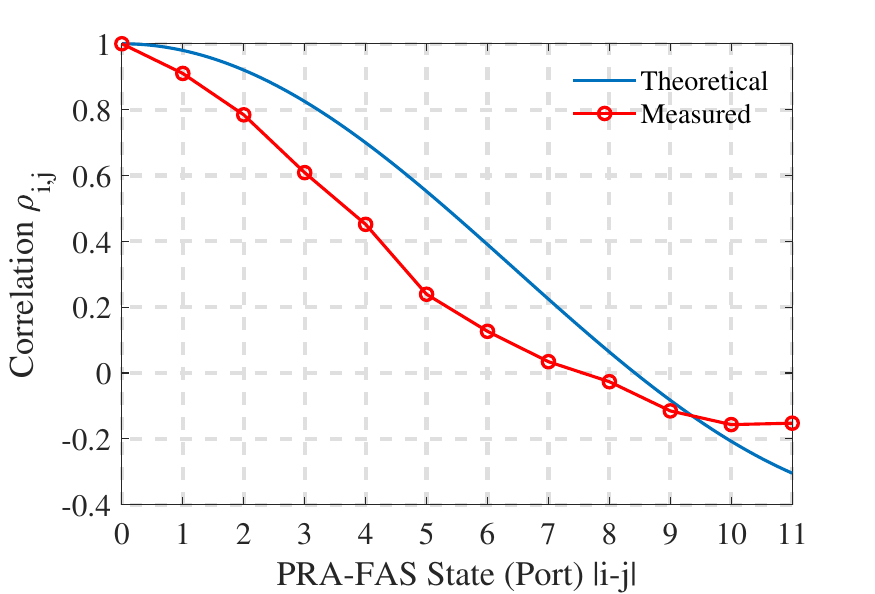}
	\caption{Theoretical and measured port correlation of the PRA-FAS.}
	\label{measured_correlation}
\end{figure}

To obtain measurements of the signal correlation between the FAS ports, we can average over multiple channel samples to provide an experimental ensemble average for the port correlation of the FAS. The measured port correlation is shown in Fig. \ref{measured_correlation}, and the correlation given by (\ref{eqn-dipole}) is also provided. It can be observed that the experimental results are generally a little less than (\ref{eqn-dipole}). This is because the rich scattering environment will not be restricted to 2D or be uniformly distributed. It practice, the PAS will be more doughnut-shaped, and therefore experimentally we will obtain correlations that are less than (\ref{eqn-dipole}). Nevertheless, the final measurement results provide concrete evidence that the PRA-FAS provides fading signals suitably correlated between FAS ports.

\section{Further Discussion}

We provide further discussion on three issues in the PRA-FAS design process. The first issue considered is the selection of the number of switches required in the PRA-FAS design. The second issue considered is explaining the reason why there is a small deviation in the covariance errors between the simulation and experimental results. The final issue considered is the bandwidth of the FAS.

\subsection{Selecting the Number of Switches}

Selecting the optimal number of RF switches is crucial in the design of the PRA-FAS to ensure a sufficient number of reconfigurable states. Striking the right balance is essential for achieving the desired flexibility in adjusting the 3D radiation pattern across all states while maintaining design simplicity and efficiency. 


To provide some insight into the selection of the number of switches, we provide results for our design algorithm for different $P$ values to obtain the corresponding average error $\delta_e$ for the geometry in  Fig. \ref{Antenna_3D}. For each $P$, the two-step method for PRA-FAS design is conducted to gain the optimal configuration $\mathbf{x}^{*}_l$, set $\mathcal{S}_k^*$, and the mapping order $\mathbf{D}^*$. The minimum average error $\delta_e$ is presented in Table \ref{Switch_quantity}. 

\begin{table}[!t]
	\caption{Average error for Various Quantity of Switches \\under the Two-Step Method for $Q=60$ and $N=12$}
	\label{Switch_quantity} 
	\centering
	\begin{tabular}{cc}
		\toprule
		Switch Quantity ($P$) & Minimum Average Error ($\delta_e$)\\
		\midrule
		4 & 0.148 \\
		5 & 0.099 \\
		6 & 0.063 \\
		7 & 0.080 \\
		8 & 0.076 \\
		\bottomrule
	\end{tabular}
\end{table}

As the number of switches $P$ increases from 4 to 6, there is a gradual reduction in the minimum average error $\delta_e$. This trend suggests that an increased number of switches allows the covariance matrix to more closely approximate the target, aligning with our initial expectations. However, upon further increasing $P$, the minimum error begins to rise. This increase is attributable to the exponential growth in the number of possible states with the increment of $P$, which means that 100 sets $\mathcal{U}^\mathrm{M}_{k,l}$ in each calculation are insufficient to explore the potential PRA-FAS configurations, thus failing to reach the global minimum in error. Taking into account the efficiency of the simulation process, we selected $P=6$ as the optimal number of switches for our design, which balances the trade-off between the complexity and performance of the PRA-FAS.

\subsection{Covariance Matrix Errors}


To help find the source of the covariance matrix error between the measured and simulated results of our PRA-FAS, it is necessary to look closely at the radiation patterns. Across all the spatial angles $\bm{\Omega}$, the amplitudes of the measured and simulated patterns were found to be consistent, however, slight differences in phase are observed. From the formula (\ref{Pattern_Correlation}), it can be deduced that phase differences are particularly influential. Even with minor variations in phase, the correlation outcomes may be seriously affected. We believe this difference in phase between the simulated and experimental results is due to the inaccuracy of the equivalent model for the RF switch in the simulation. Because we use 6 switches, any small errors will be accumulated even further. One approach to rectifying the switch modeling issue is to perform an optimization of the equivalent switch model parameters, so that the difference between the model and experimental results are matched.


\subsection{Bandwidth}

The focus of this work has been to demonstrate that FAS can be constructed using PRA. The bandwidth of the prototype obtained is just over 50 MHz. In wireless applications, wider bandwidths will be required. Approaches to increasing the bandwidth are therefore an important issue to consider. In principle, this can be achieved by increasing $T$ in (\ref{GA2_updated}) however there are two constraints that will also need to be overcome. The first is due to the design of the PRA-FAS radiating feed E-slot patch. Its height from the ground plane is low at only 1.524 mm resulting in the restricted bandwidth. The second is the limit of using only 6 switches for the reconfiguration of the pixel surface. 

To obtain a wider bandwidth for the FAS to handle stable correlations for a larger bandwidth, it is necessary to increase the number of switches. That is we need to search through more switch combinations to find those with sufficient bandwidth. We have found that doubling the height of the radiating feed plate and increasing the number of switches to 7 can allow us to increase the bandwidth to 130 MHz, exceeding 5\%. Future investigation of increasing the bandwidth is required so that other approaches can also be considered for bandwidth extension. 

\section{Conclusion}

In this paper, we have described a new approach to FAS antenna design based on PRA. In developing the approach, we show that fluid antennas can be considered as being equivalent to antennas with $''$fluid$''$ radiation patterns. To validate the design approach, simulation and experimental results for a PRA-FAS prototype controlled by RF switches for a FAS with $W=0.5$ and $N=12$ operating at a center frequency of 2.5 GHz are provided. Because the PRA-FAS uses electronic switches for reconfiguration, it can meet the required packet-by-packet switching speeds of FAS. 
 
It is experimentally shown that the 12 states of the PRA-FAS can provide channels that vary significantly when the underlying wireless channel is stationary, demonstrating that the PRA-FAS can provide the necessary diversity. Additionally, it is shown that the pattern covariance matrix approximately meets Clarke's model both by simulation and experiment. Within the PRA-FAS, 6 RF switches are strategically placed among the 60 internal ports, striking a balance between complexity and performance. To realize this PRA-FAS's intricate design, we employed a two-step optimization process, including a random search and GA optimization. This method sequentially refines the antenna configuration and the sequence of matching patterns, aiming to approximate the ideal covariance matrix. 

Further research is required. Methods to extend the PRA size for larger FAS scales with $W>\lambda/2$ and $N>12$ are required to be developed, and multi-port PRA-FAS designs also need to be investigated for MIMO-FAS. 

\bibliographystyle{IEEEtran}
\bibliography{IEEEabrv, IEEEreference}

\vfill

\end{document}